\documentclass{aa}  

\usepackage{graphicx}
\usepackage{txfonts}
\usepackage{hyperref}
\usepackage{url}
\usepackage{fontawesome}

\def\Nlens{23}

\def\mchab{M_*^{\mathrm{(Chab)}}}
\def\mchabi{M_{*,i}^{\mathrm{(Chab)}}}

\def\mhalo{M_{h}}

\def\chalo{c_{h}}

\def\reff{R_{\mathrm{e}}}

\def\rein{R_{\mathrm{Ein}}}
\def\mein{M_{\mathrm{Ein}}}
\def\tein{\theta_{\mathrm{Ein}}}
\def\teini{\theta_{\mathrm{Ein},i}}

\def\nser{n}
\def\aimf{\alpha_{\mathrm{IMF}}}

\def\hyperp{\boldsymbol{\eta}}
\def\individ{\boldsymbol{\psi}}
\def\individi{\boldsymbol{\psi}_i}

\def\data{\mathbf{d}}
\def\datai{\mathbf{d}_i}

\def\crosssect{\sigma_{\mathrm{SL}}}

\def\Sref#1{Sect.~\ref{#1}\xspace}
\def\Fref#1{Fig.~\ref{#1}\xspace}
\def\Tref#1{Table~\ref{#1}\xspace}
\def\Eref#1{Eq.~\ref{#1}\xspace}

\def\pr{{\rm P}}
\def\fsel{\mathcal{F}_{\mathrm{sel}}}
\def\fdet{\mathcal{F}_{\mathrm{det}}}
\def\pcmass{{\rm P}_{\mathrm{CMASS}}}

\begin{document}

   \title{Survey of gravitationally-lensed objects in HSC imaging (SuGOHI). III. Statistical strong lensing constraints on the stellar IMF of CMASS galaxies.}
   \titlerunning{SuGOHI III. Stellar IMF and dark matter of CMASS galaxies.}
   \authorrunning{Sonnenfeld et al.}

   \author{Alessandro Sonnenfeld\inst{1,2}\thanks{Marie Sk\l{}odowska-Curie Fellow}\and
          Anton T. Jaelani\inst{3,4}\and
          James Chan\inst{5}\and
          Anupreeta More\inst{1,6}\and
          Sherry H. Suyu\inst{7,8,9}\and
          Kenneth C. Wong\inst{2,10}\and
          Masamune Oguri\inst{2,11,12}\and
          Chien-Hsiu Lee\inst{13}
          }

   \institute{Leiden Observatory, Leiden University, Niels Bohrweg 2, 2333 CA Leiden, the Netherlands\\
              \email{sonnenfeld@strw.leidenuniv.nl}
             \and
            Kavli IPMU (WPI), UTIAS, The University of Tokyo, Kashiwa, Chiba 277-8583, Japan\and
            Faculty of Science and Engineering, Kindai University, Higashi-Osaka 577-8502, Japan\and
            Astronomical Institute, Tohoku University, Aramaki, Aoba, Sendai 980-8578, Japan\and
            Institute of Physics, Laboratory of Astrophysics, \'{E}cole Polytechnique F\'{e}d\'{e}rale de Lausanne (EPFL), Observatoire de Sauverny, 1290 Versoix, Switzerland\and
            The Inter-University Center for Astronomy and Astrophysics, Post bag 4, Ganeshkhind, Pune, 411007, India\and
            Max-Planck-Institut f\"ur Astrophysik, Karl-Schwarzschild Str. 1, 85741 Garching, Germany\and
            Physik-Department, Technische Universit\"at M\"unchen, James-Franck-Stra\ss{}e~1, 85748 Garching, Germany\and
            Institute of Astronomy and Astrophysics, Academia Sinica, 11F of ASMAB, No.1, Section 4, Roosevelt Road, Taipei 10617, Taiwan\and
            National Astronomical Observatory of Japan, 2-21-1 Osawa, Mitaka, Tokyo 181-8588, Japan\and
            Department of Physics, University of Tokyo, 7-3-1 Hongo, Bunkyo-ku, Tokyo 113-0033, Japan\and
            Research Center for the Early Universe, University of Tokyo, 7-3-1 Hongo, Bunkyo-ku, Tokyo 113-0033, Japan\and
            National Optical Astronomy Observatory, 950 N Cherry Ave, Tucson, AZ 85719, USA
             }

   \date{}

 
  \abstract
   {The determination of the stellar initial mass function (IMF) of massive galaxies is one of the open problems in cosmology. Strong gravitational lensing is one of the few methods that allow us to constrain the IMF outside of the Local Group.}
   {The goal of this study is to statistically constrain the distribution in the IMF mismatch parameter, defined as the ratio between the true stellar mass of a galaxy and that inferred assuming a reference IMF, of massive galaxies from the Baryon Oscillation Spectroscopic Survey (BOSS) constant mass (CMASS) sample.} 
   {
We took $\Nlens$ strong lenses drawn from the CMASS sample, measured their Einstein radii and stellar masses using multi-band photometry from the Hyper Suprime-Cam survey, then fitted a model distribution for the IMF mismatch parameter and dark matter halo mass to the whole sample. We used a prior on halo mass from weak lensing measurements and accounted for strong lensing selection effects in our model.
}
   {
Assuming a Navarro Frenk \& White density profile for the dark matter distribution, we infer a value $\mu_{\mathrm{IMF}} = -0.04\pm0.11$ for the average base-10 logarithm of the IMF mismatch parameter, defined with respect to a Chabrier IMF.
A Salpeter IMF is in tension with our measurements.}
   {
Our results are consistent with a scenario in which the region of massive galaxies where the IMF normalisation is significantly heavier than that of the Milky Way is much smaller than the scales $5-10$~kpc probed by the Einstein radius of the lenses in our sample, as recent spatially-resolved studies of the IMF in massive galaxies suggest.
The Monte Carlo chains describing the posterior probability distribution of the model are available online \href{https://github.com/astrosonnen/sugohi3_inference}{\faGithub}, together with the code used to obtain them.
}


   \maketitle
%

\section{Introduction}\label{sect:intro}

The stellar initial mass function (IMF) is one of the fundamental properties of stellar populations.
Within the Milky Way, the stellar IMF is observed to be a relatively constant function across a wide range of environments \citep[see e.g.][]{BCM10}.
For other galaxies, however, it is difficult to obtain direct (i.e. based on star counts) measurements of the IMF, especially outside the Local Group.

The question of whether the stellar IMF is a universal function or not is an important one: detecting a variation, or lack thereof, of the IMF with galaxy properties can give us insight into the physics of star formation.
Moreover, the vast majority of the measurements of stellar masses of galaxies at cosmological distances used in the literature rely on the assumption of a particular form of the IMF. If the IMF is not universal, these measurements are biased. Not knowing the IMF limits our ability to match observations with theoretical predictions: for instance, varying the IMF will shift the stellar mass function and the stellar mass-size relation, two distributions that are commonly used to assess the accuracy of hydrodynamical cosmological simulations.

The past decade has seen a relatively large number of observational studies aimed at constraining the stellar IMF in massive early-type galaxies (ETGs). Many of these studies suggest that the IMF of these objects is different from that observed in the Milky Way, resulting in a higher stellar mass-to-light ratio at fixed age and metallicity. These include works based on the analysis of IMF-sensitive absorption lines in integrated spectra of massive galaxies \citep{v+C10, Spi++12, Fer++13}, on the combination of strong lensing and stellar dynamics \citep{Aug++10, Bar++13, Son++15}, on dynamical modelling of nearby galaxies with integrated field unit spectroscopic data \citep{Cap++12, Lyu++16, Li++17}, and on gravitational microlensing of strongly lensed quasars \citep{Sch++14}.
There are, however, massive ETGs known for having an IMF similar to that of the Milky Way, in terms of mass-to-light ratio \citep{S+L13, SLC15, CSL18}.

From the theoretical point of view, a number of models have been proposed to explain the observed shape of the IMF and its variation across the galaxy population \citep[see][and references therein]{Kru14}. 
However, as recently shown by \citet{Gus++19}, finding models that can simultaneously reproduce the near-universality of the IMF in the Milky Way and the bottom-heavy IMFs suggested by observations of massive ETGs is very challenging.

The picture is complicated by the possible presence of radial gradients in the IMF, detected in some spatially resolved studies of IMF-sensitive absorption features \citep[though some other studies find contrasting results. See][for a complete picture]{Mar++15, LaB++16, van++17, Sar++18, Par++18,Zie++17,ASL17,Vau++18}, suggested by observations of mass-to-light ratio gradients from lensing and dynamics \citep{SLE17b, O+A18a, O+A18b, Son++18b, Col++18}, and predicted in cosmological simulations with a non-universal IMF \citep{BSC18c}.
With gradients in the IMF, more care is required when comparing observations with models, since the interpretation of a given measurement will depend on the spatial scale over which it is carried out.

Strong gravitational lensing is one of the few available methods for constraining the stellar IMF in objects outside of the Local Group. Strong lensing can provide a very precise measurement of the total projected mass enclosed within the Einstein radius of a galaxy, the radius enclosing an average surface mass density equal to the lensing critical density, which typically probes scales around $5-10$~kpc from the centre.
This mass measurement can be converted into a mass-to-light ratio, which can then be compared to IMF-dependent predictions from stellar population synthesis modelling.

The presence of dark matter, however, complicates the interpretation of strong lensing measurements. 
Typically, the degeneracy between the dark matter mass enclosed within the Einstein radius and the stellar mass-to-light ratio ($M_*/L$ from now on) is broken by combining lensing with other probes, such as stellar dynamics \citep{Tre++10}, by statistically combining a large set of lenses \citep{ORF14}, or by combining both approaches \citep{Son++15}.
In this work, the third paper of the Survey of Gravitationally-lensed Objects in Hyper Suprime-Cam (HSC) Imaging \citep[SuGOHI]{Son++18a,Won++18}, we use strong lensing measurements in combination with weak lensing information to constrain the IMF of a statistical sample of galaxies at $z\sim0.6$.

We focus on massive galaxies drawn from the constant mass (CMASS) subset of the Baryon Oscillation Spectroscopic Survey \citep[BOSS][]{SWE09, Daw++13} in the Sloan Digital Sky Survey III \citep[SDSS-III][]{Eis++11}.
We use photometric data from the HSC \citep{Miy++18} Subaru Strategic Program \citep{Aih++18a} to measure Einstein radii and stellar masses of $\Nlens$ CMASS strong lenses.
This sample includes lenses presented in \citet[][, Paper I from now on]{Son++18a} and \citet[][Paper II from now on]{Won++18}, as well as lens systems from the literature that are covered by the HSC survey.

We then take advantage of the recent weak lensing study of \citet[][S19 from here on]{Son++19}, who used HSC data to constrain the stellar-to-halo mass relation (SHMR) of CMASS galaxies, to put a prior on the halo mass distribution of our strong lens sample and thus break the degeneracy between the dark matter distribution and the stellar IMF.
An important caveat is that our CMASS strong lenses, although drawn from the general population of CMASS galaxies, are not a representative sample of the latter, due to selection effects: galaxies with a larger lensing cross-section are more likely to be lenses, and lenses with different properties have different probabilities of being detected by a lensing survey.
We explicitly take these effects into account as part of our model. 
This is one of the most important features of our method, which enables us to use information obtained from the CMASS sample as a whole as a prior on the strong lens sample. 

For all CMASS galaxies, central stellar velocity dispersion measurements are available from BOSS spectroscopy. In principle, we could use these measurements as additional constraints on the gravitational potential of the lens. We choose not to for two reasons: firstly, BOSS velocity dispersion measurements of CMASS galaxies are very noisy; secondly, carrying out a joint lensing and stellar dynamics study with single aperture velocity dispersion measurements requires making a series of additional assumptions on the geometry and the orbital structure of each lens \citep[typically spherical symmetry, isotropic orbits, and a spatially constant $M_*/L$. See e.g.][]{Aug++10, Son++15}. 
Some of these assumptions might bias the results \citep[see][for discussions on the effect of assuming a spatially constant $M_*/L$ on the stellar IMF inferred from lensing and dynamics]{Son++18b, Ber++18}.
We decide instead to use purely lensing data to constrain our model. 
As a result, our observational constraints on individual lenses are rather limited, and the ability to statistically combine our measurements in a meaningful and accurate way plays a central role in our inference.
To emphasise these features, we label our method statistical strong lensing.

The structure of this work is as follows.
In \Sref{sect:data} we describe the sample of lenses and the data used for this study.
In \Sref{sect:models} we perform lens modelling and a stellar population synthesis analysis to obtain measurements of the Einstein radius and estimates of the stellar mass of each lens.
In \Sref{sect:method} we introduce the model distribution used to fit the population of lenses. 
We first carry out our analysis with a simplified version of the model. Then, in \Sref{sect:deteff}, we repeat the analysis with the full model.
We discuss our findings in \Sref{sect:discuss} and summarise our results in \Sref{sect:concl}.
We assume a flat $\Lambda$CDM cosmology with $\Omega_M=0.3$ and $H_0=70\,\rm{km}\,\rm{s}^{-1}\,\rm{Mpc}^{-1}$. Magnitudes are in AB units. All images are oriented with north up and east left.


\section{Data}\label{sect:data}

\subsection{Lens sample}\label{ssec:sample}

Our starting sample consists of strong lenses from the CMASS sample of BOSS galaxies for which HSC imaging data in $grizy$ bands is available.
As of the S17A internal data release of the HSC survey, there are 84 between definite and probable lenses (grade A and B, using the notation of Paper I) that satisfy this requirement. Seventy-two of these were recently discovered using HSC data \citep[see][Paper I and Paper II for details]{Tan++16}. Nine belong to the Strong Lensing Legacy Survey \citep[SL2S][]{Ruf++11, Gav++12, Mor++12, Son++13a}, while three are part of the BOSS Emission-Line Lens Survey \citep[BELLS][]{Bro++12}.
Hyper Suprime-Cam and SL2S lenses have been selected by means of searches for arc-like images of strongly lensed sources in photometric data, while the selection of the BELLS sample is based on a search for emission lines from strongly lensed objects in the BOSS spectra of luminous red galaxies.

In order to carry out our strong lensing analysis, we require systems for which the redshift of both the lens and the background source are known.
For all CMASS lenses, the spectroscopic redshift of the lens galaxy is available from the BOSS catalogue.
Of the 84 lenses from the starting sample, 16 have spectroscopic redshifts of the source from the literature.
We measured source redshifts for an additional seven systems, thanks to a spectroscopic follow-up campaign on the Very Large Telescope (VLT), the details of which are given in Sect. \ref{ssec:xshoo}.

Finally, for a more straightforward modelling and interpretation of the strong lensing data, we require lenses to consist of only one massive deflector, excluding systems with two or more galaxies of comparable mass acting as lenses.
Although this requirement is already satisfied by all the lens systems in our sample for which we have source redshifts, it is important to take this condition into account when discussing selection effects (see Sect. \ref{sect:deteff}).
Our final sample then consists of $\Nlens$ CMASS strong gravitational lenses with HSC imaging and spectroscopic redshifts of both the lens and the source. These lenses are listed in \Tref{tab:coords} with their coordinates, lens and source redshifts, lens stellar velocity dispersion as obtained from the SDSS data release 12 \citep[DR12][]{Ala++15}, and references. 
\begin{table*}
\caption{Name, coordinates, lens redshift, source redshift, lens stellar velocity dispersion (from SDSS DR12), and references for the sample of strong lenses used for our study. All lens galaxies belong to the CMASS sample of BOSS.}
\label{tab:coords}
\begin{tabular}{lcccccc}
\hline
\hline
Name & R.A. & Dec & $z_d$ & $z_s$ & $\sigma_{\mathrm{BOSS}}$ & References\\
 & (deg) & (deg) & & & (km s$^{-1}$) & \\
\hline
HSCJ015618$-$010747 & 29.07554 & -1.12977 & 0.542 & 1.167 & $299\pm38$ & Paper II \\
HSCJ020241$-$064611 & 30.67247 & -6.76979 & 0.502 & 2.748 & $162\pm25$ & Paper I \\
HSCJ021411$-$040502 & 33.54670 & -4.08411 & 0.609 & 1.880 & $196\pm104$ & \citet{Ruf++11,Gav++12} \\
 & & & & & & \citet{Mor++12}\\ 
HSCJ021737$-$051329 & 34.40492 & -5.22482 & 0.646 & 1.847 & $271\pm47$ & \citet{Ruf++11,Gav++12} \\
HSCJ022346$-$053418 & 35.94227 & -5.57180 & 0.499 & 1.444 & $259\pm26$ & \citet{Son++13a,Son++13b} \\
HSCJ022610$-$042011 & 36.54440 & -4.33656 & 0.496 & 1.232 & $226\pm38$ & \citet{Son++13a,Son++13b} \\
HSCJ023307$-$043838 & 38.27945 & -4.64396 & 0.671 & 1.869 & $408\pm107$ & \citet{Mor++12,Son++15} \\
HSCJ023817$-$054555 & 39.57402 & -5.76542 & 0.599 & 1.763 & $291\pm71$ & Paper I \\
HSCJ085855$-$010208 & 134.73330 & -1.03567 & 0.468 & 1.421 & $199\pm32$ & Paper I \\
HSCJ094427$-$014742 & 146.11446 & -1.79511 & 0.539 & 1.179 & $179\pm38$ & \citet{Bro++12} \\
HSCJ120623$+$001507 & 181.59937 & 0.25199 & 0.563 & 3.120 & $273\pm46$ & Paper I \\
HSCJ121052$-$011905 & 182.71869 & -1.31810 & 0.700 & 2.295 & $347\pm78$ & Paper I \\
HSCJ121504$+$004726 & 183.76850 & 0.79056 & 0.642 & 1.297 & $709\pm86$ & \citet{Bro++12} \\
HSCJ140929$-$011410 & 212.37381 & -1.23631 & 0.584 & 2.302 & $191\pm38$ & Paper I \\
HSCJ141300$-$012608 & 213.25030 & -1.43560 & 0.749 & 2.666 & $373\pm107$ & Paper I \\
HSCJ141815$+$015832 & 214.56556 & 1.97564 & 0.556 & 2.139 & $199\pm26$ & Paper I \\
HSCJ142449$-$005321 & 216.20420 & -0.88934 & 0.795 & 1.302 & $295\pm55$ & \citet{Tan++16} \\
HSCJ142720$+$001916 & 216.83562 & 0.32114 & 0.551 & 1.266 & $240\pm33$ & Paper I \\
HSCJ144307$-$004056 & 220.77985 & -0.68225 & 0.500 & 1.071 & $274\pm34$ & Paper I \\
HSCJ220506$+$014703 & 331.27884 & 1.78441 & 0.476 & 2.526 & $285\pm46$ & \citet{Mor++12,Son++13a,Son++13b} \\
HSCJ222801$+$012805 & 337.00824 & 1.46826 & 0.647 & 2.462 & $349\pm56$ & Paper I \\
HSCJ223733$+$005015 & 339.38973 & 0.83772 & 0.604 & 2.143 & $226\pm50$ & Paper I \\
HSCJ230335$+$003703 & 345.89654 & 0.61755 & 0.458 & 0.936 & $266\pm38$ & \citet{Bro++12} \\

\end{tabular}
\end{table*}
The typical velocity dispersion of our lenses is in the range $200-300$~km~s$^{-1}$, with some outliers, including a galaxy with a nominal value of $\sigma_{\mathrm{BOSS}} = 709\pm86$~km~s$^{-1}$. These are most likely the result of systematic effects in the measurements of the stellar velocity dispersion, related to the low signal-to-noise ratio of BOSS spectra.

\subsection{Hyper Suprime-Cam photometry}

\begin{figure*}
\begin{tabular}{cc}
\includegraphics[width=0.47\textwidth]{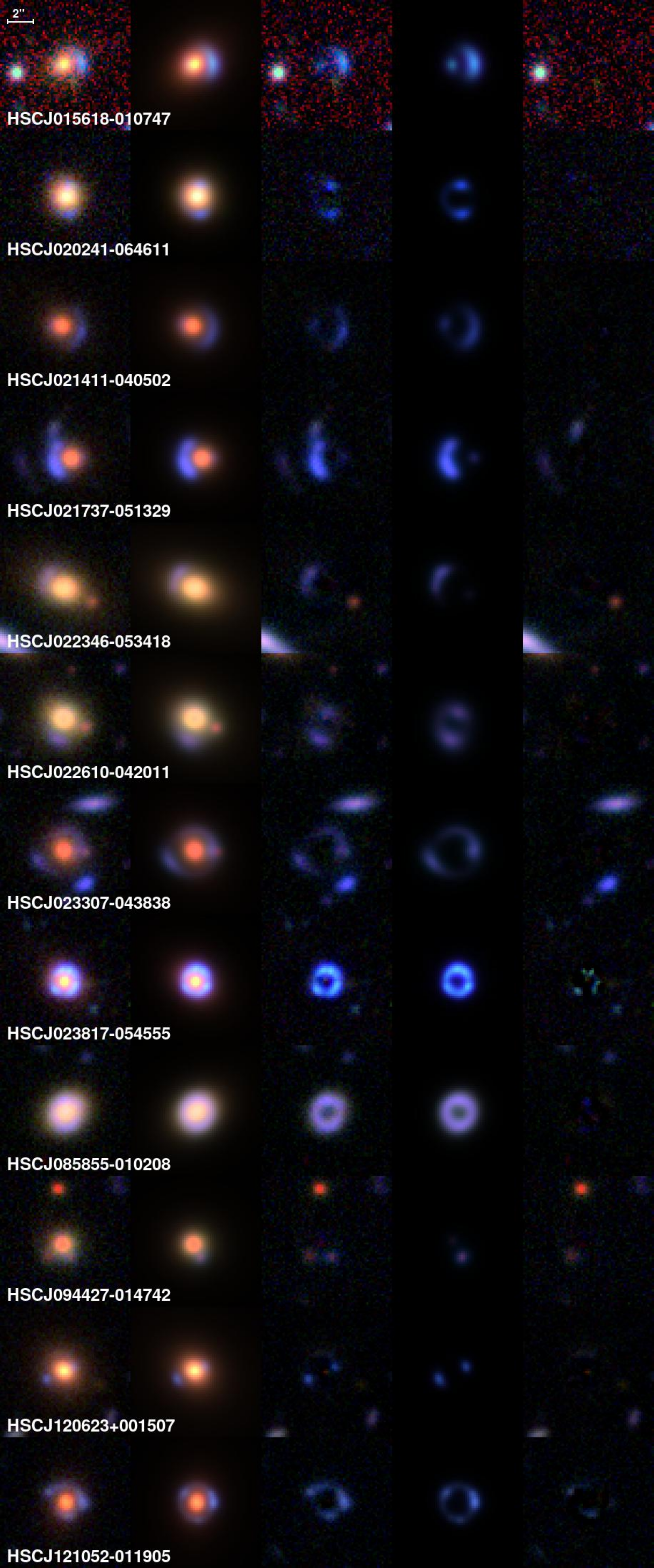} &
\includegraphics[width=0.47\textwidth]{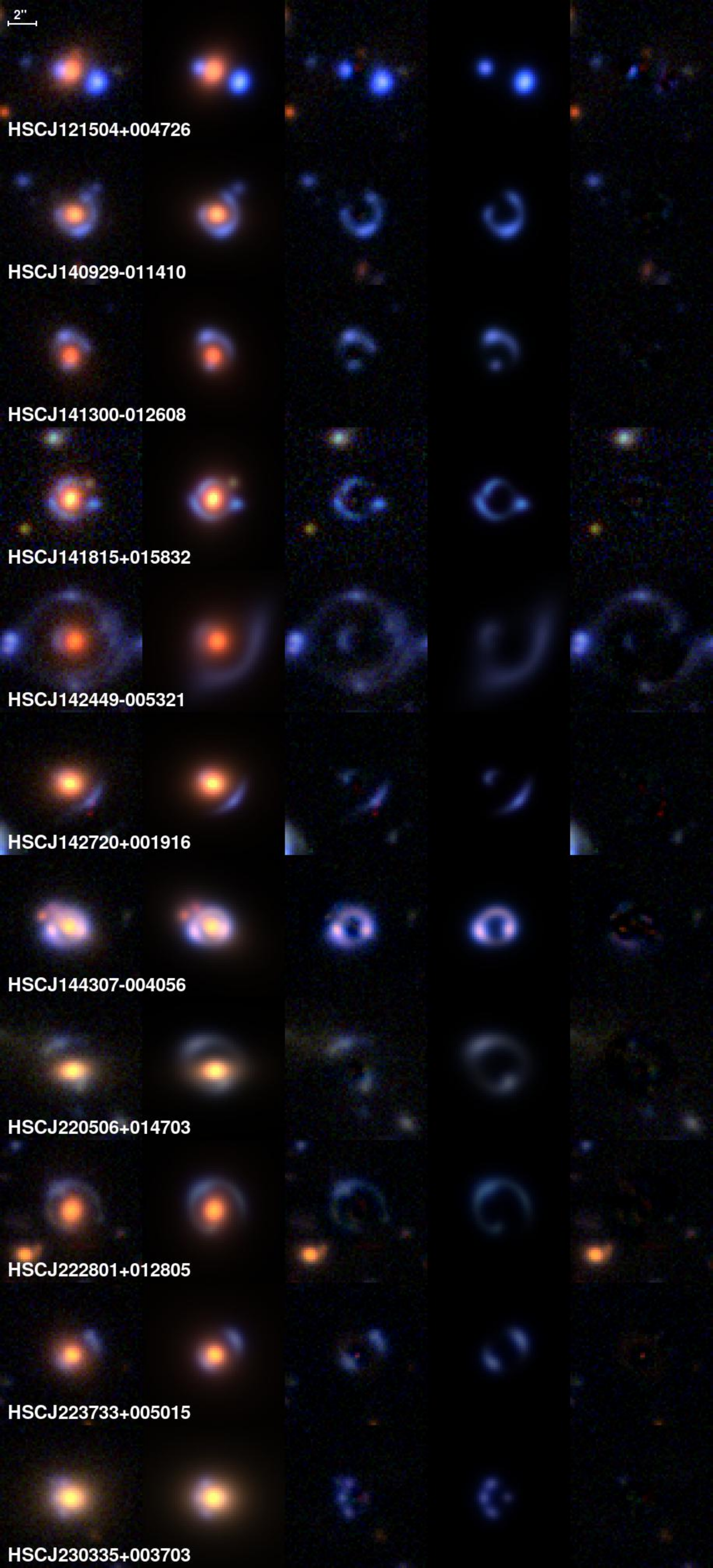}
\end{tabular}
\caption{ Twenty-three CMASS galaxy strong lenses in our sample. First column: Colour-composite HSC images in $g,r,i$ bands. Second column: Best-fit lens + source model. Third column: Data with the best-fit model of the lens light subtracted. Fourth column: Dest-fit source model. Fifth column: Residuals.}
\label{fig:lensmodels}
\end{figure*}

We use photometric data from the S17A internal release of the HSC survey to obtain stellar mass and Einstein radius measurements of the lens galaxies in our sample. 
Data from S17A  has been processed with the HSC data reduction pipeline {\sc HSCPipe} version 5.4 \citep{Bos++18}, a version of the Large Synoptic Survey Telescope pipeline \citep{Ive++08, Axe++10, Jur++15}.
The end products of {\sc HSCPipe} include sky-subtracted coadded images, variance maps, and models of the point spread function (PSF), which we use for our analysis.
In particular, in each of the $g,r,i,z,y$ bands, we obtain $101\times101$~pixel cutouts ($16.4''\times16.4''$) centred on the lens, as well as samples of the PSF on a $37\times37$ grid, with the same pixel size as the data.
Colour-composite images in $g,r,i$ bands of the $\Nlens$ lenses subject of our study are shown in the left column of \Fref{fig:lensmodels}.

\subsection{Very Large Telescope spectroscopy}\label{ssec:xshoo}

We observed nine CMASS lens candidates from the SuGOHI sample (paper I) with the X-Shooter spectrograph \citep{Ver++11} on the VLT (ESO programme 099.A-0220, PI Suyu), with the main goal of measuring the redshift of the lensed background source.
Each target was observed in slit mode during either one or two observation blocks (OBs), depending on the brightness of the source. Each OB corresponds to roughly one hour of telescope time, and consists of $10\times285\rm{s}$ exposures obtained in an ABBA nodding pattern, to optimise background subtraction in the near-infrared (NIR) arm.
Exposure times in the ultraviolet (UVB) and visible (VIS) arms are slightly shorter due to the longer readout time.
We used slit widths of $1.0''$, $0.9'',$ and $0.9''$ in the UVB, VIS, and NIR arms, respectively, and applied a $2\times2$ pixel binning to the UVB and VIS CCDs. We positioned the slit so that it covered both the centre of the lens galaxy and the brightest feature of the lensed source.
Observations were executed with a seeing full width at half maximum $\rm{FWHM} < 0.9''$ on target position.

We used 2D spectra for our analysis, as provided by the ESO Quality Control Group, obtained by processing the data with the X-Shooter pipeline \citep{Mod++10}.
We visually inspected the spectra of each system, looking for emission lines from the source. For seven of the nine observed lens candidates we see multiple emission lines, which enable us to measure the source redshift. No emission lines are visible in the spectrum of HSCJ140705$-$011256 and HSCJ142053+005620, therefore we do not use these systems for our lensing study.
 
\begin{figure}
\includegraphics[width=\columnwidth]{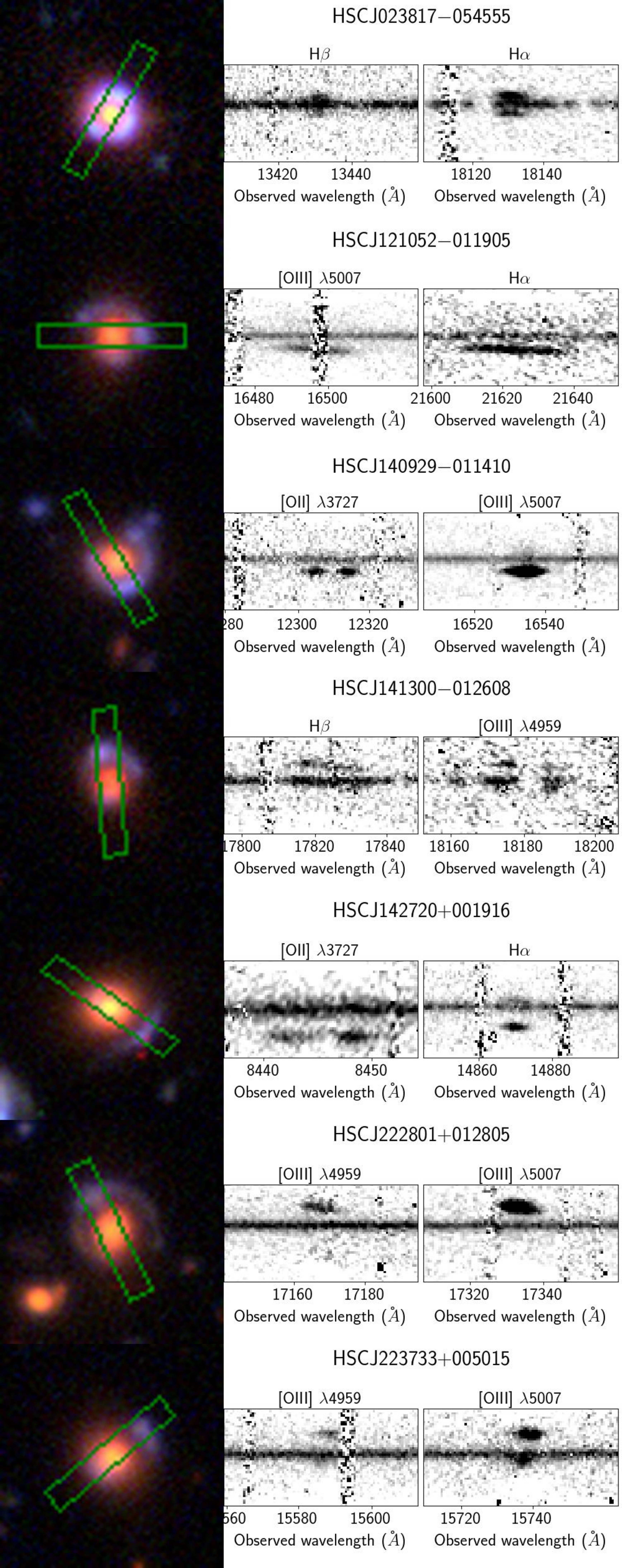}
\caption{X-Shooter observations of seven CMASS lenses with detected emission lines from the lensed source. The left panel shows an HSC $gri$ colour-composite image of the lens. The green box indicates the position of the slit used during the spectroscopic observation. The middle and right panels show small cutouts of the 2D spectrum around the two source emission lines with the highest signal-to-noise ratio.}
\label{fig:xshoo}
\end{figure}

We summarise the spectroscopic observations in \Tref{tab:xshoo}, where, in the last column, we list the emission lines from the lensed source that are detected in the spectrum of each system. Small cutouts of the 2D spectrum of each lens around the two emission lines with the highest signal-to-noise ratio are shown in \Fref{fig:xshoo}, together with a colour-composite image of the lens and a box indicating the position of the slit. Only lenses with detected emission lines are shown.
Three of the lenses in this sample, HSCJ142720$+$001916, HSCJ222801$+$012805, and HSCJ223733$+$005015, were classified as grade B lens candidates in Paper I, meaning that their lens nature could not be determined with certainty with the available data, which consisted only of HSC photometry. 
X-Shooter observations confirmed that the blue arcs are indeed at a higher redshift compared to the main galaxy, and therefore lensed by it. This additional piece of information, together with the fact that we are able to reproduce the observed image configuration with a simple model, as we will show in Sect. \ref{ssec:lensmodels}, allows us to upgrade these candidates to grade A lenses.

\begin{table*}
\caption{Spectroscopic observations of nine CMASS lens candidates from the SuGOHI sample. Each row corresponds to one observation block. The third column indicates the position angle of the slit, in degrees east of north. The fourth and fifth columns list the source redshift and the emission lines detected in the spectrum.}
\label{tab:xshoo}
\begin{tabular}{lcccl}
\hline
\hline
Lens name & Obs. date & P.A. & $z_s$ & Em. lines \\
\hline
HSCJ023817$-$054555 & 2017-11-18 & -30 & 1.763 & H$\alpha$, H$\beta$, [OII]$\lambda$3727\\
 & 2017-11-16 & -30 &  & \\
HSCJ121052$-$011905 & 2017-04-05 & 90 & 2.295 & H$\alpha$, [OIII]$\lambda$5007, [OII]$\lambda$3727\\
HSCJ140705$-$011256 & 2017-07-27 & 125 & $\cdots$ & None\\
HSCJ140929$-$011410 & 2018-03-01 & 30 & 2.302 & H$\alpha$, [OIII]$\lambda$5007, [OIII]$\lambda$4959, H$\beta$, [OII]$\lambda$3727\\
HSCJ141300$-$012608 & 2017-04-17 & 5 & 2.666 & H$\beta$, [OIII]$\lambda$4959\\
HSCJ142053$+$005620 & 2017-04-07 & 150 & $\cdots$ & None\\
HSCJ142720$+$001916 & 2017-04-08 & 55 & 1.266 & H$\alpha$, [OII]$\lambda$3727\\
HSCJ222801$+$012805 & 2017-07-25 & 25 & 2.462 & [OIII]$\lambda$5007, [OIII]$\lambda$4959, H$\beta$, Ly$\alpha$\\
 & 2017-08-05 & 25 &  & \\
HSCJ223733$+$005015 & 2017-07-08 & -50 & 2.143 & H$\alpha$, [OIII]$\lambda$5007, [OIII]$\lambda$4959, [OII]$\lambda$3727\\
 & 2017-07-08 & -50 &  & \\

\end{tabular}
\end{table*}

\section{Lens models and stellar mass measurements}\label{sect:models}

We wish to measure the Einstein radius and the stellar mass, or, more specifically, the stellar mass density profile, of the $\Nlens$ lenses in our sample. 
Our strategy consists of 1) fitting a parameterised model for the lens and source galaxy system to $grizy$ HSC images, 2) fitting a stellar population synthesis model to the lens galaxy fluxes measured in step 1.
These two steps are described separately in the next two subsections.

\subsection{Lens modelling}\label{ssec:lensmodels}

Our model consists of two light components, describing the lens and source galaxy respectively, and one mass component associated with the lens.
The surface brightness distribution of the lens is described with a S\'{e}rsic profile \citep{Ser68} with elliptical isophotes, while the source is modelled as an exponential profile (i.e. a S\'{e}rsic profile with index $\nser=1$), also elliptical.
For each component, we assume the structural parameters (the centroid, half-light radius, axis ratio, position angle, and S\'{e}rsic index) to have the same values in all bands, allowing only the total flux to vary with wavelength. This corresponds to a model with spatially constant colours.

We model the lens mass with a singular isothermal ellipsoid \citep[SIE,][]{KSB94}. Although the true density profiles of the lenses in our sample may in general be different from isothermal, the measurement of the Einstein radius is robust to the particular choice of the lens model to a few percent accuracy \citep{Bol++08}.
We impose the centroid of the mass to be coincident with that of the lens light component, but allow the axis ratio and position angle to be different.

For each system, we search for the set of values of the model parameters that minimises the $\chi^2$ between the seeing-convolved model surface brightness distribution and the observed data in each band. For the fit, we use a circular region extending out to a radius where the lens surface brightness falls below the level of the sky fluctuations (typically between $3''$ and $5''$). Objects not associated with either the lens or the source are masked out or explicitly modelled as an extra S\'{e}rsic component.

We explore the parameter space by running a Markov Chain Monte Carlo (MCMC), using the software {\sc emcee} \citep{For++13}. 
At each step of the chain, we draw a set of values of the structural parameters of the lens and source light, of the lens mass (Einstein radius, axis ratio, and position angle), as well as the four colours of the lens and the source, defined with respect to the $i-$band. We then run an optimiser to find the lens and source $i-$band magnitudes that minimise the $\chi^2$.
We assume a flat prior on all model parameters.

In \Fref{fig:lensmodels}, we show, for each lens, colour-composite $gri$ images of the observed data, the best-fit model, the data with the best-fit model of the lens light subtracted, the best-fit model of the source galaxy only, and the residual image.
In \Tref{tab:lightmodel} we report the best-fit values of the parameters describing the surface brightness distribution of the lens galaxy, while the parameters describing the lens mass and the source surface brightness are listed in \Tref{tab:lensmodel}.
We define the Einstein radius as the circularised radius of the elliptical isodensity curve that encloses an average surface mass density equal to the lensing critical density.
\begin{table*}
\caption{Best-fit values of the lens light S\'{e}rsic profile model parameters, including the half-light radius, S\'{e}rsic index, axis ratio, position angle of the major axis (east of north), and magnitudes in $g,r,i,z,y$ bands. 
The last row lists the median $1\sigma$ statistical uncertainties on each parameter.}
\label{tab:lightmodel}
\begin{tabular}{lccccccccc}
\hline
\hline
Name & $\reff$ & $\nser$ & $b/a$ & PA & $m_{l,g}$ & $m_{l,r}$ & $m_{l,i}$ & $m_{l,z}$ & $m_{l,y}$ \\
    & $('')$ & & & (deg) & & & & & \\
\hline
HSCJ015618$-$010747 & $3.36$ & $8.44$ & $0.667$ & $132.2$ & $21.78$ & $19.89$ & $18.76$ & $18.40$ & $18.19$ \\
HSCJ020241$-$064611 & $0.68$ & $2.78$ & $0.938$ & $9.9$ & $21.69$ & $20.07$ & $19.21$ & $18.84$ & $18.68$ \\
HSCJ021411$-$040502 & $3.37$ & $8.38$ & $0.931$ & $52.8$ & $22.34$ & $20.53$ & $19.30$ & $18.80$ & $18.59$ \\
HSCJ021737$-$051329 & $1.26$ & $7.16$ & $0.953$ & $82.6$ & $22.42$ & $20.79$ & $19.59$ & $19.11$ & $18.85$ \\
HSCJ022346$-$053418 & $2.17$ & $6.88$ & $0.637$ & $69.3$ & $21.07$ & $19.37$ & $18.47$ & $18.06$ & $17.89$ \\
HSCJ022610$-$042011 & $1.71$ & $7.89$ & $0.786$ & $57.5$ & $21.33$ & $19.54$ & $18.64$ & $18.23$ & $18.05$ \\
HSCJ023307$-$043838 & $1.55$ & $5.56$ & $0.801$ & $25.0$ & $22.26$ & $20.70$ & $19.43$ & $18.97$ & $18.72$ \\
HSCJ023817$-$054555 & $0.71$ & $4.59$ & $0.777$ & $46.4$ & $21.97$ & $20.30$ & $19.18$ & $18.71$ & $18.53$ \\
HSCJ085855$-$010208 & $0.95$ & $4.29$ & $0.715$ & $123.6$ & $21.77$ & $20.05$ & $19.19$ & $18.79$ & $18.58$ \\
HSCJ094427$-$014742 & $1.14$ & $6.43$ & $0.930$ & $85.5$ & $22.21$ & $20.55$ & $19.57$ & $19.14$ & $18.95$ \\
HSCJ120623$+$001507 & $3.36$ & $9.26$ & $0.841$ & $44.6$ & $21.37$ & $19.74$ & $18.68$ & $18.26$ & $18.05$ \\
HSCJ121052$-$011905 & $2.20$ & $7.78$ & $0.725$ & $171.6$ & $22.34$ & $20.70$ & $19.39$ & $18.93$ & $18.66$ \\
HSCJ121504$+$004726 & $0.72$ & $4.24$ & $0.711$ & $151.2$ & $22.32$ & $20.80$ & $19.55$ & $19.07$ & $18.80$ \\
HSCJ140929$-$011410 & $1.23$ & $5.79$ & $0.749$ & $82.2$ & $22.31$ & $20.72$ & $19.60$ & $19.15$ & $18.96$ \\
HSCJ141300$-$012608 & $1.60$ & $6.03$ & $0.772$ & $27.4$ & $22.84$ & $21.19$ & $19.80$ & $19.30$ & $19.01$ \\
HSCJ141815$+$015832 & $0.95$ & $5.20$ & $0.737$ & $172.0$ & $22.04$ & $20.41$ & $19.35$ & $18.92$ & $18.72$ \\
HSCJ142449$-$005321 & $2.53$ & $4.68$ & $0.932$ & $-20.1$ & $22.45$ & $20.63$ & $19.21$ & $18.56$ & $18.26$ \\
HSCJ142720$+$001916 & $0.89$ & $4.53$ & $0.825$ & $94.6$ & $21.97$ & $20.27$ & $19.17$ & $18.71$ & $18.53$ \\
HSCJ144307$-$004056 & $0.75$ & $4.32$ & $0.614$ & $60.8$ & $21.82$ & $20.20$ & $19.25$ & $18.81$ & $18.62$ \\
HSCJ220506$+$014703 & $0.59$ & $4.77$ & $0.463$ & $92.7$ & $22.01$ & $20.35$ & $19.47$ & $19.06$ & $18.85$ \\
HSCJ222801$+$012805 & $1.11$ & $3.74$ & $0.791$ & $168.2$ & $22.31$ & $20.66$ & $19.47$ & $18.99$ & $18.72$ \\
HSCJ223733$+$005015 & $3.05$ & $9.82$ & $0.956$ & $55.4$ & $21.81$ & $20.17$ & $19.00$ & $18.50$ & $18.29$ \\
HSCJ230335$+$003703 & $1.64$ & $5.06$ & $0.783$ & $85.5$ & $21.26$ & $19.56$ & $18.70$ & $18.32$ & $18.11$ \\
\hline 
Typical uncertainties & $0.12$ & $0.06$ & $0.004$ & $0.5$ & $0.01$ & $0.01$ & $0.01$ & $0.01$ & $0.01$ \\

\end{tabular}
\end{table*}

The median S\'{e}rsic index of the lens galaxies is $5.6$, with four galaxies with $n > 8$. 
While these are relatively large values of the S\'{e}rsic index, compared to the canonical picture of quiescent galaxies being described by $n=4$ models, this distribution is very similar to that measured by S19 on CMASS galaxies.
Although errors on the S\'{e}rsic index propagate into measurements of the total flux and the half-light radius of a galaxy, we point out that, for the purposes of constraining the stellar IMF with strong lensing, it is sufficient to obtain an accurate description of the stellar profile in the region enclosed by the Einstein radius. This is very well constrained by HSC photometry and robust to changes in the surface brightness profile.
The estimate of the total stellar mass of a galaxy only matters for the purpose of assigning a prior on the dark matter halo mass, using the SHMR measured by S19. Since both the data and the analysis method we employ here are the same as those used by S19, this procedure is insensitive to possible systematics in the surface brightness profile fitting process.

The Einstein radius is, in most cases, well constrained by HSC data, owing to the presence of extended arcs and/or counter-images detected with high signal-to-noise ratio. The only exception is HSCJ094427$-$014742, a lens from the BELLS sample: for this system, only one image of the source is visible, and is not sufficient to obtain a robust lens model.
We then fix the lens model parameters to the values measured by \citet{Bro++12} with Hubble Space Telescope (HST) data, in which the counter-image is visible.

One of the lenses in the sample is the double source plane lens HSCJ142449-005321, the `Eye of Horus' lens \citep{Tan++16}. For consistency with the rest of the sample, we only model the lensing effect on the source closer to us, forming the inner ring, while masking out the light from the outer ring. The value of the Einstein radius reported in \Tref{tab:lensmodel} then refers to a source redshift of $z_s = 1.302$.

The values of the typical statistical uncertainty on each parameter, defined as the median across the sample of the 68\% enclosed probability interval, as obtained from the MCMC, are given at the bottom of \Tref{tab:lightmodel} and \Tref{tab:lensmodel}.
These are typically very small, thanks to the depth of HSC imaging data.
In practice, however, systematic uncertainties related to the choice of model are larger than statistical ones.
\citet{Bol++08} have estimated systematic uncertainties on the Einstein radius to be on the order of a few percent, when high resolution HST imaging data is used to constrain $\tein$. In our case, we are using lower resolution ground-based data, which could lead to larger systematic errors on the lens model.
In addition to HSCJ094427$-$014742, models obtained using HST data are available from the literature for three more lenses in our sample: SL2S lenses HSCJ021411$-$040502 and HSCJ021737$-$051329, and BELLS lens HSCJ230335$+$003703. We can use these lenses to get a rough estimate of the robustness of our HSC-based measurements of $\tein$.
\citet{Son++13a} measured $\tein=1.41''$ for HSCJ021411$-$040502 and $\tein=1.27''$ for HSCJ021737$-$051329, while \citet{Bro++12} measured $\tein=1.01''$ for HSCJ230335$+$003703. Our estimates are respectively 11\% smaller, 2\% smaller, and 2\% larger than their values.
Given the outcome of this comparison, we assume a 10\% systematic uncertainty on $\tein$ for all the lenses in our sample.

\begin{table*}
\caption{Best-fit values of the lens mass and source surface brightness model parameters, including the Einstein radius, lens mass axis ratio, position angle of the major axis of the lens mass (east of north), source-plane (i.e. de-lensed) half-light radius, distance between the source and lens centroid along the right ascension and declination direction, and magnitudes in $g,r,i,z,y$ bands of the background source. The last row lists the median $1\sigma$ statistical uncertainties on each parameter.}
\label{tab:lensmodel}
\begin{tabular}{lccccccccccc}
\hline
\hline
Name & $\tein$ & $b/a$ & PA & $R_{e,\mathrm{source}}$ & $\Delta$ R.A. & $\Delta$ Dec & $m_{s,g}$ & $m_{s,r}$ & $m_{s,i}$ & $m_{s,z}$ & $m_{s,y}$ \\
  & $('')$ & & (deg) & $('')$ & & & & & \\
  & $('')$ & & (deg) & $('')$ & $('')$ & $('')$ & & & & & \\
\hline
HSCJ015618$-$010747 & $0.84$ & $0.99$ & $174.3$ & $0.27$ & $-0.519$ & $-0.041$ & $23.41$ & $23.12$ & $23.05$ & $22.30$ & $22.28$ \\
HSCJ020241$-$064611 & $1.16$ & $0.85$ & $127.9$ & $0.10$ & $-0.000$ & $0.258$ & $25.35$ & $25.27$ & $25.58$ & $25.66$ & $26.10$ \\
HSCJ021411$-$040502 & $1.25$ & $0.80$ & $76.3$ & $0.27$ & $-0.395$ & $0.134$ & $24.86$ & $24.68$ & $24.76$ & $24.79$ & $24.90$ \\
HSCJ021737$-$051329 & $1.25$ & $0.60$ & $95.6$ & $0.07$ & $0.288$ & $-0.001$ & $24.94$ & $24.86$ & $24.81$ & $24.69$ & $24.67$ \\
HSCJ022346$-$053418 & $1.47$ & $0.76$ & $64.0$ & $0.07$ & $0.271$ & $-0.164$ & $26.58$ & $26.18$ & $25.99$ & $25.64$ & $25.27$ \\
HSCJ022610$-$042011 & $1.04$ & $0.70$ & $77.3$ & $0.46$ & $0.174$ & $0.599$ & $23.76$ & $23.36$ & $23.05$ & $22.56$ & $22.65$ \\
HSCJ023307$-$043838 & $1.72$ & $0.83$ & $51.5$ & $0.21$ & $0.357$ & $0.092$ & $24.94$ & $24.47$ & $24.51$ & $24.32$ & $24.24$ \\
HSCJ023817$-$054555 & $0.93$ & $0.92$ & $119.8$ & $0.08$ & $-0.030$ & $-0.014$ & $24.93$ & $24.87$ & $24.96$ & $25.10$ & $25.09$ \\
HSCJ085855$-$010208 & $1.01$ & $0.93$ & $118.7$ & $0.28$ & $-0.029$ & $0.071$ & $24.01$ & $23.36$ & $22.88$ & $22.45$ & $22.16$ \\
HSCJ094427$-$014742 & $0.72$ & $0.92$ & $108.0$ & $0.06$ & $-0.197$ & $0.353$ & $25.42$ & $25.01$ & $25.05$ & $24.35$ & $24.17$ \\
HSCJ120623$+$001507 & $1.16$ & $0.94$ & $41.7$ & $0.08$ & $0.316$ & $0.149$ & $25.93$ & $25.89$ & $26.04$ & $25.99$ & $26.05$ \\
HSCJ121052$-$011905 & $1.28$ & $0.91$ & $67.4$ & $0.05$ & $-0.011$ & $0.031$ & $27.02$ & $26.73$ & $26.73$ & $26.51$ & $26.46$ \\
HSCJ121504$+$004726 & $1.52$ & $0.91$ & $112.4$ & $0.05$ & $-0.525$ & $0.171$ & $22.86$ & $22.78$ & $22.81$ & $22.36$ & $23.12$ \\
HSCJ140929$-$011410 & $1.24$ & $0.70$ & $68.6$ & $0.12$ & $-0.139$ & $-0.065$ & $25.23$ & $24.97$ & $25.02$ & $25.00$ & $24.90$ \\
HSCJ141300$-$012608 & $1.13$ & $0.67$ & $-26.5$ & $0.13$ & $-0.027$ & $-0.376$ & $24.72$ & $24.26$ & $24.19$ & $24.11$ & $24.02$ \\
HSCJ141815$+$015832 & $1.39$ & $0.68$ & $175.4$ & $0.13$ & $-0.253$ & $0.108$ & $24.90$ & $24.65$ & $24.49$ & $24.36$ & $24.33$ \\
HSCJ142449$-$005321 & $2.14$ & $0.58$ & $33.9$ & $0.95$ & $-0.704$ & $0.258$ & $23.37$ & $22.96$ & $22.65$ & $22.05$ & $21.90$ \\
HSCJ142720$+$001916 & $1.40$ & $0.88$ & $94.9$ & $0.14$ & $-0.487$ & $0.300$ & $24.75$ & $24.60$ & $24.31$ & $24.04$ & $23.98$ \\
HSCJ144307$-$004056 & $1.10$ & $0.82$ & $46.8$ & $0.12$ & $0.247$ & $0.007$ & $23.90$ & $23.37$ & $22.96$ & $22.73$ & $22.57$ \\
HSCJ220506$+$014703 & $1.70$ & $0.99$ & $103.0$ & $0.43$ & $0.391$ & $-0.553$ & $24.24$ & $23.49$ & $23.23$ & $23.04$ & $23.05$ \\
HSCJ222801$+$012805 & $1.73$ & $0.61$ & $1.9$ & $0.24$ & $-0.110$ & $-0.266$ & $25.36$ & $25.00$ & $25.20$ & $25.30$ & $25.27$ \\
HSCJ223733$+$005015 & $1.29$ & $0.79$ & $27.5$ & $0.23$ & $-0.405$ & $-0.359$ & $24.57$ & $24.36$ & $24.28$ & $24.17$ & $24.09$ \\
HSCJ230335$+$003703 & $1.03$ & $0.44$ & $91.1$ & $0.08$ & $0.207$ & $0.014$ & $25.39$ & $25.36$ & $25.21$ & $25.42$ & $24.02$ \\
\hline 
Typical uncertainties & $0.02$ & $0.01$ & $0.8$ & $0.03$ & $0.004$ & $0.004$ & $0.02$ & $0.03$ & $0.02$ & $0.04$ & $0.05$ \\

\end{tabular}
\end{table*}

\subsection{Stellar population synthesis}

We measure the stellar mass of the lens galaxies by fitting composite stellar population (CSP) models to the observed $g,r,i,z,y$ magnitudes, following a Bayesian procedure similar to that used by \citet{Aug++09}.
We use the stellar population synthesis code {\sc BC03} \citep{B+C03} to generate CSP models with an exponentially decaying star formation history and a Chabrier IMF \citep{Cha03}. In particular, we obtain model spectra over a grid of values of the following parameters: age (i.e. time since the first burst of star formation), star formation decay time, metallicity, and dust attenuation.
Then, for each lens galaxy, we calculate the predicted flux in each HSC filter at each point of the grid.

We sample the parameter space defined by the stellar mass and the other parameters of the CSP model by running an MCMC. We assume flat priors on age, star formation decay time, on the logarithm of the dust attenuation parameter, and on the logarithm of the stellar mass, while we assume a Gaussian prior on the logarithm of the metallicity at fixed stellar mass, following the observational study by \citet{Gal++05}.
We obtain the model magnitudes at any point within the grid by interpolation.
We correct the observed magnitudes for galactic extinction using the dust reddening map of \citet{S+F11}, as provided by the NASA Infrared Science Archive.

\begin{figure}
\includegraphics[width=\columnwidth]{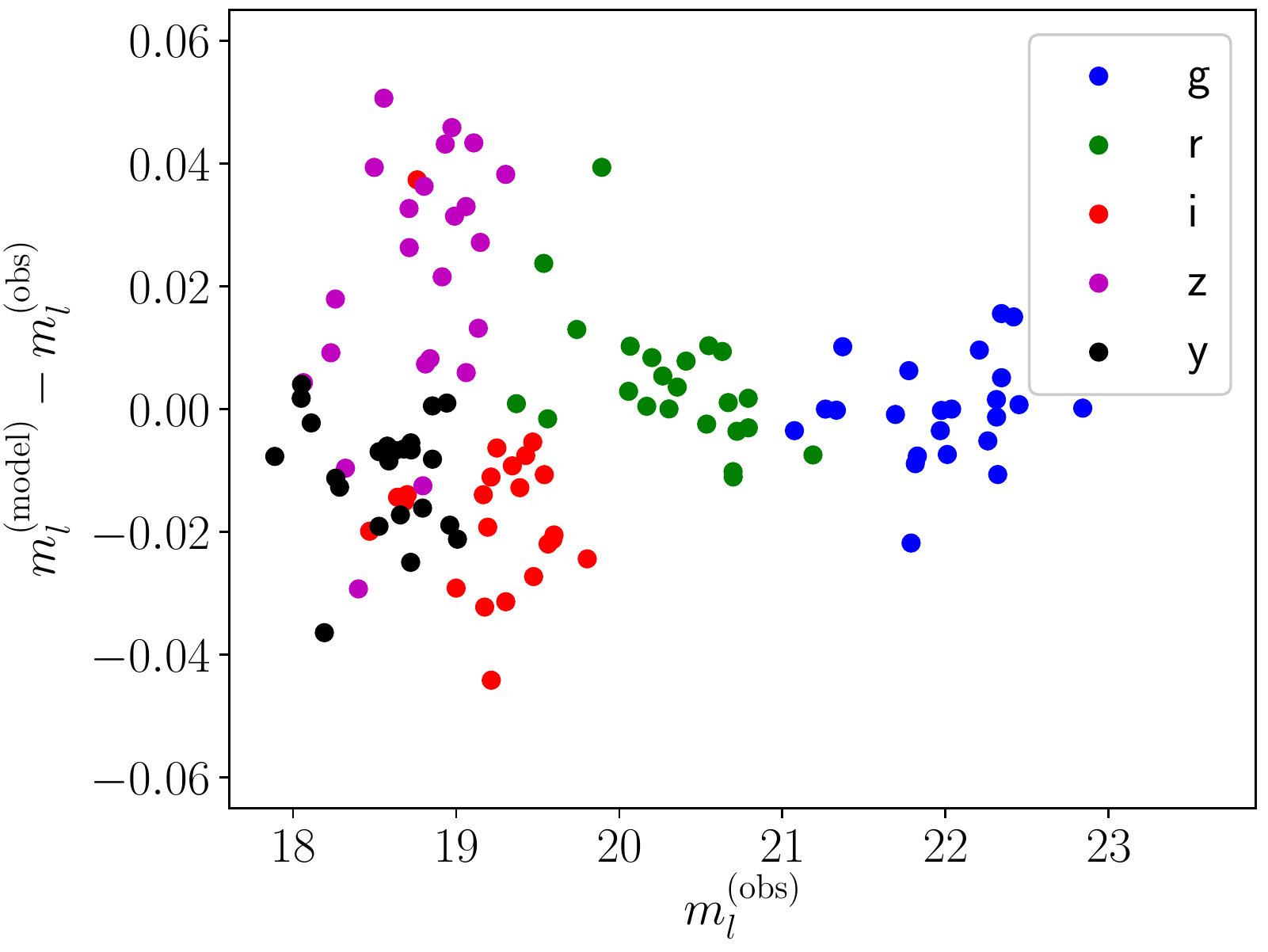}
\caption{Difference between magnitudes predicted by the CSP model that maximises the likelihood and the observed magnitudes, as a function of the latter. Different colours correspond to different HSC filters. The typical observational uncertainty on the observed magnitudes, from S\'{e}rsic profile fitting, is $0.02$.}
\label{fig:sedmismatch}
\end{figure}

Typical uncertainties on the observed fluxes are $\sim0.02$ magnitudes. Our CSP models are unable to fit such precise data down to the noise level, despite having the same number of degrees of freedom (five) as the number of data points, for each galaxy. 
This is shown in \Fref{fig:sedmismatch}, where we plot the difference between the observed and predicted magnitudes for the CSP model that maximises the likelihood.
As can be seen, the deviations are often larger than the statistical uncertainty on the observed magnitudes, especially for the redder bands.
In order to obtain a reliable estimate of the uncertainty on the stellar mass, following S19 we add in quadrature a $0.05$~mag uncertainty to the data, which is the typical standard deviation between the data and the magnitudes predicted by the best-fit CSP model. This procedure allows us to take into account, to some extent, systematic errors associated with the model.

The inferred values of the stellar mass, with their $1\sigma$ uncertainty, are listed in \Tref{tab:mass}. Throughout this paper, we indicate the stellar mass obtained from stellar population synthesis modelling as $\mchab$, to highlight the fact that this measurement relies on the assumption of a Chabrier IMF.
Also in \Tref{tab:mass}, we list the value of the Einstein radius in physical units of each lens, $\rein$, the corresponding enclosed total projected mass, $\mein$, and the stellar mass enclosed within $\rein$, obtained assuming a constant $M_*/L$ throughout the galaxy.

\begin{figure}
\includegraphics[width=\columnwidth]{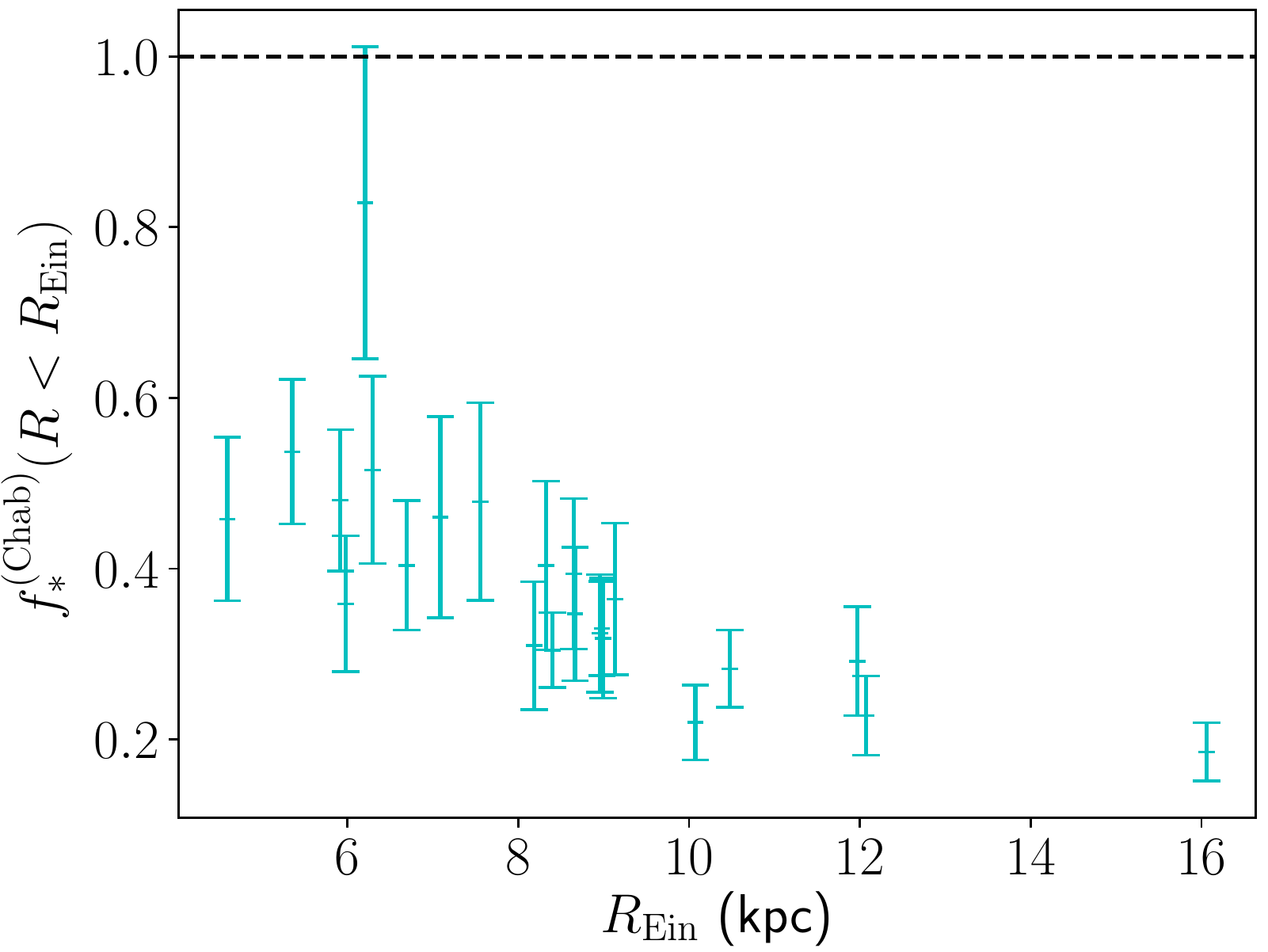}
\caption{Ratio between the stellar mass enclosed within the Einstein radius, as inferred from stellar population synthesis modelling assuming a Chabrier IMF, and the total mass within the Einstein radius, as a function of the Einstein radius. The horizontal dashed line corresponds to the limit of 100\% mass fraction within the Einstein radius. Values above this limit are non-physical.}
\label{fig:fchab}
\end{figure}

Finally, in \Fref{fig:fchab}, we plot the derived fraction of the mass enclosed within the Einstein radius that is accounted for by stellar mass, as derived from stellar population synthesis modelling. The lenses with the largest Einstein radii have a comparatively smaller stellar mass fraction, meaning that dark matter dominates the mass within the Einstein radius for these systems.
In particular, HSCJ142449$-$005321 is the lens with both the largest Einstein radius and the smallest stellar mass fraction. As first pointed out by \citet{Tan++16}, this lens is identified as the most probable brightest galaxy of a rich ($\sim50$ members) cluster, according to the HSC cluster catalogue of \citet{Ogu++18}.
One of the lenses, HSCJ023817$-$054555, appears to have a significantly larger value of the stellar mass fraction, compared to the rest of the sample. 
This lens has simultaneously one of the smallest half-light radii and Einstein radii of the whole sample: $\reff=0.71''$ and $\tein=0.93''$.
The large value of $f_*$ is then due to the fact that strong lensing is probing the inner regions of a compact galaxy, which we expect to be dominated by stars.
\begin{table*}
\caption{Mass measurements on the strong lenses in our sample. From left to right these are stellar mass inferred from stellar population synthesis fitting of HSC photometry, Einstein radius in physical units, total projected mass enclosed within the Einstein radius, and stellar population synthesis stellar mass enclosed within an aperture equal to the Einstein radius.}
\label{tab:mass}
\begin{tabular}{lcccc}
\hline
\hline
Name & $\log{\mchab}$ & $\rein$ & $\log{\mein}$ & $\log{\mchab(<\rein)}$ \\
  & $\log{M_\odot}$ & (kpc) & $\log{M_\odot}$ & $\log{M_\odot}$ \\
\hline
015618$-$010747 & $11.72 \pm 0.07$ & $5.36$ & $11.40$ & $11.13 \pm 0.07$ \\
020241$-$064611 & $11.31 \pm 0.11$ & $7.09$ & $11.48$ & $11.14 \pm 0.11$ \\
021411$-$040502 & $11.66 \pm 0.06$ & $8.40$ & $11.68$ & $11.16 \pm 0.06$ \\
021737$-$051329 & $11.56 \pm 0.10$ & $8.67$ & $11.72$ & $11.26 \pm 0.10$ \\
022346$-$053418 & $11.67 \pm 0.10$ & $8.96$ & $11.78$ & $11.29 \pm 0.10$ \\
022610$-$042011 & $11.62 \pm 0.09$ & $6.30$ & $11.51$ & $11.22 \pm 0.09$ \\
023307$-$043838 & $11.65 \pm 0.09$ & $12.07$ & $12.01$ & $11.37 \pm 0.09$ \\
023817$-$054555 & $11.59 \pm 0.10$ & $6.21$ & $11.43$ & $11.35 \pm 0.10$ \\
085855$-$010208 & $11.39 \pm 0.08$ & $5.92$ & $11.42$ & $11.10 \pm 0.08$ \\
094427$-$014742 & $11.32 \pm 0.10$ & $4.60$ & $11.26$ & $10.93 \pm 0.10$ \\
120623$+$001507 & $11.70 \pm 0.11$ & $7.55$ & $11.52$ & $11.20 \pm 0.11$ \\
121052$-$011905 & $11.70 \pm 0.11$ & $9.14$ & $11.73$ & $11.29 \pm 0.11$ \\
121504$+$004726 & $11.60 \pm 0.07$ & $10.48$ & $11.99$ & $11.44 \pm 0.07$ \\
140929$-$011410 & $11.41 \pm 0.11$ & $8.19$ & $11.62$ & $11.11 \pm 0.11$ \\
141300$-$012608 & $11.61 \pm 0.11$ & $8.32$ & $11.63$ & $11.24 \pm 0.11$ \\
141815$+$015832 & $11.44 \pm 0.10$ & $8.99$ & $11.71$ & $11.22 \pm 0.10$ \\
142449$-$005321 & $12.07 \pm 0.08$ & $16.06$ & $12.46$ & $11.73 \pm 0.08$ \\
142720$+$001916 & $11.55 \pm 0.07$ & $8.98$ & $11.83$ & $11.35 \pm 0.07$ \\
144307$-$004056 & $11.44 \pm 0.08$ & $6.69$ & $11.61$ & $11.22 \pm 0.08$ \\
220506$+$014703 & $11.26 \pm 0.09$ & $10.07$ & $11.80$ & $11.14 \pm 0.09$ \\
222801$+$012805 & $11.61 \pm 0.10$ & $11.97$ & $11.95$ & $11.41 \pm 0.10$ \\
223733$+$005015 & $11.73 \pm 0.10$ & $8.65$ & $11.68$ & $11.28 \pm 0.10$ \\
230335$+$003703 & $11.51 \pm 0.10$ & $5.98$ & $11.55$ & $11.10 \pm 0.10$ \\

\end{tabular}
\end{table*}

\section{Population analysis}\label{sect:method}

In this section, we carry out a Bayesian hierarchical inference of the distribution of the stellar IMF, as well as other galaxy properties entering the problem, of CMASS galaxies, given the strong lensing and stellar population synthesis measurements presented in the previous section.

\subsection{Individual object parameters}\label{ssec:model}

Let us introduce the IMF mismatch parameter, $\aimf$, defined as the ratio between the true stellar mass of a galaxy and the value of the stellar mass inferred assuming a Chabrier IMF and having an otherwise perfect knowledge of the remaining stellar population parameters:
\begin{equation}\label{eq:aimfdef}
\aimf \equiv \frac{M_*^{\mathrm{(true)}}}{\mchab}.
\end{equation}
We wish to infer the distribution of $\aimf$ of CMASS galaxies, given the measurements of $\rein$ and $\mchab$ on our $\Nlens$ strong lenses. 

In order for us to evaluate the likelihood of these measurements, we need to propose a model for the mass distribution of the lenses in our sample, with $\aimf$ being one of the model parameters.
We describe each lens as the sum of a stellar and a dark matter mass component.
We assume that the stellar surface mass density follows a S\'{e}rsic profile, with half-light radius $\reff$ and S\'{e}rsic index $\nser$ fixed to the values inferred from the fit of \Sref{sect:models} and total mass $M_* = \mchab\aimf$.
We are implicitly assuming that the stellar population parameters, including the stellar IMF, are constant throughout a given galaxy.
We will discuss the implications of this assumption for our inference, in light of existing evidence in favour of gradients in the stellar IMF, in Sect. \ref{ssec:assumpt}.
We then assume that the dark matter density follows a Navarro Frenk \& White profile \citep[NFW,][]{NFW97},
\begin{equation}
\rho(r) \propto \frac{1}{r(1+r/r_s)^2}.
\end{equation}
We define the halo mass as the mass enclosed within a sphere with average density equal to 200 times the critical density of the Universe. 
We then describe the profile in terms of the concentration parameter, defined as the ratio between the radius of the shell enclosing a mass equal to $M_h$, $r_{200}$, and the scale radius $r_s$:
\begin{equation}
c_h \equiv \frac{r_{200}}{r_s}.
\end{equation}

For a gravitational lens, the value of the Einstein radius depends not only on the mass distribution of the lens, but also on the geometry of the system, that is, the redshift of the lens and the source, $z_d$ and $z_s$.
We can then fully describe a strong lens system with the set of parameters
\begin{equation}\label{eq:individ}
\individ \equiv \{\mchab,\aimf,\nser,\reff,\mhalo,\chalo,z_d,z_s\},
\end{equation}
to which we will often refer collectively as `individual object parameters' (as opposed to parameters describing the whole population) and which we summarise with the symbol $\individ$.

It is not possible to constrain the exact values of $\individ$ for each lens, given our data: the Einstein radius determines only the total enclosed mass, so that the same value of $\tein$ can be reproduced with different relative amounts of stellar and dark matter.
Our goal, however, is to infer how the individual object parameters are distributed across the population of galaxies, with particular attention paid to the IMF mismatch parameter $\aimf$.
For this purpose, we introduce a population distribution described by a set of hyper-parameters $\hyperp$, which can be interpreted as a prior on the individual lens parameters:
\begin{equation}
\pr(\individ) = \pr(\individ|\hyperp). \nonumber
\end{equation}
Our goal is to infer the posterior probability distribution of the hyper-parameters given the data, $\pr(\hyperp|\data)$.

\subsection{Population distribution}

Our strong lenses are drawn from the CMASS sample of galaxies. We then describe the population distribution of the individual object parameters of the lenses as the product between the distribution of the whole CMASS sample, $\pcmass(\individ|\hyperp)$, and a selection function term $\fsel(\individ|\hyperp)$, which takes into account the fact that some objects (e.g. more massive galaxies) are more likely to be strong lenses, depending on the value of $\individ$:
\begin{equation}\label{eq:psplit}
\pr(\individ|\hyperp) = \fsel(\individ|\hyperp) \pcmass(\individ|\hyperp).
\end{equation}
In this subsection, we focus on the term $\pcmass$, while we discuss the selection function term $\fsel$ in Sect. \ref{ssec:selfunc}.
Strictly speaking, $\pcmass$ is not a probability distribution: it is not normalised to unity, only the product $\fsel\pcmass$ is. Nonetheless, it can be considered as such when studied on its own.

We assume that $\pcmass$ factorises as
\begin{equation}\label{eq:cmassterms}
\begin{split}
\pcmass(\individ|\hyperp) = & \mathcal{S}(\mchab|\hyperp)\mathcal{N}(\nser|\mchab,\hyperp)\mathcal{R}(\reff|\mchab,\nser,\hyperp) \times \\
& \mathcal{H}(\mhalo|\mchab,\hyperp)\mathcal{C}(\chalo|\mhalo,\hyperp)\mathcal{A}(\aimf,\hyperp) \times \\
& \mathcal{Z}_d(z_d|\hyperp)\mathcal{Z}_s(z_s|\hyperp).
\end{split}
\end{equation}

The product $\mathcal{S}\mathcal{N}\mathcal{R}$ describes the distribution of stellar mass, S\'{e}rsic index, and half-light radius of CMASS galaxies. 
Following S19, we assume
$\mathcal{S}$ to be a skew-Gaussian distribution in $\log{\mchab}$, $\mathcal{N}$ a Gaussian distribution in $\log{\nser}$ with mean that scales linearly with $\log{\mchab}$ , and $\mathcal{R}$ a Gaussian distribution in $\log{\reff}$ with mean that scales linearly with $log{\mchab}$ and $\log{\nser}$. We refer to Sect. 3.2 of S19 for a detailed description of this part of the model.
A total of ten hyper-parameters describes these three terms.
These hyper-parameters have been measured with a high precision by S19, owing to their large sample size of $\sim10,000$ CMASS galaxies. 
In order to reduce the dimensionality of the problem and simplify calculations, we fix them to the values reported in Table 2 of S19.

The term $\mathcal{H}$ describes the distribution in halo mass of CMASS galaxies.
Again, following S19, we assume this to be a log-Gaussian distribution,
\begin{equation}\label{eq:mhalodist}
\mathcal{H}(\mhalo) = \frac{1}{\sqrt{2\pi}\sigma_h} \exp{\left\{-\frac{(\log{\mhalo} - \mu_h(\mchab))^2}{2\sigma_h^2}\right\}}.
\end{equation}
In their analysis, S19 let the mean of this Gaussian scale with stellar mass and half-light radius. However, they did not find any evidence for a dependence of halo mass on $\reff$. 
In light of their results, we assume the following form for the mean of $\log{\mhalo}$:
\begin{equation}\label{eq:muhalo}
\mu_h(\mchab) = \mu_{h,0} + \beta_h(\log{\mchab} - 11.4),
\end{equation}
where $\beta_h$ is the slope of the power-law relation between halo and stellar mass, and $\mu_{h,0}$ is the average value of $\log{\mhalo}$ at the pivot stellar mass $\log{\mchab}=11.4$.
The term $\mathcal{C}$ describes the halo mass-concentration relation, which we describe as a Gaussian in $\log{\chalo}$,
\begin{equation}\label{eq:chalodist}
\mathcal{C}(\chalo) = \frac{1}{\sqrt{2\pi}\sigma_c} \exp{\left\{-\frac{(\log{\chalo} - \mu_c(\mhalo))^2}{2\sigma_c^2}\right\}},
\end{equation}
with mean
\begin{equation}\label{eq:muchalo}
\mu_c(\mhalo) = 0.830 - 0.098 (\log{\mhalo} - 12)
\end{equation}
and dispersion $\sigma_c=0.1$.
The values of the coefficients of the above equation are the same used by S19 and are taken from the \citet{Mac++08} study.
We are implicitly assuming that the mass-concentration relation is independent of redshift. 
This is a reasonable approximation, because the predicted change in concentration over the narrow redshift range spanned by our lenses is smaller than the intrinsic scatter around the mean relation \citep{Mac++08}.

The distribution in the IMF mismatch parameter is described by the term $\mathcal{A}$, which we model as a Gaussian in $\log{\aimf}$,
\begin{equation}\label{eq:aimfdist}
\mathcal{A}(\aimf) = \frac{1}{\sqrt{2\pi}\sigma_{\mathrm{IMF}}} \exp{\left\{-\frac{(\log{\aimf} - \mu_{\mathrm{IMF}})^2}{2\sigma_{\mathrm{IMF}}^2}\right\}},
\end{equation}
with mean $\mu_{\mathrm{IMF}}$ and dispersion $\sigma_{\mathrm{IMF}}$. These are the main parameters of interest in our study.

The term $\mathcal{Z}_d$ is the redshift distribution of CMASS galaxies, which we approximate as a Gaussian:
\begin{equation}
\mathcal{Z}_d(z_d|\hyperp) = \frac{1}{\sqrt{2\pi}\sigma_d}\exp{\left\{-\frac{(z_d - \mu_d)^2}{2\sigma_d^2}\right\}}.
\end{equation}
Fitting this distribution to the full set of CMASS galaxies, we infer $\mu_d=0.558$ and $\sigma_d=0.085$ with very small errors.
We then keep these hyper-parameters fixed, as done for the hyper-parameters describing the distribution in stellar mass, S\'{e}rsic index, and half-light radius. 

Finally, $\mathcal{Z}_s$ describes the distribution of source redshifts, which must be interpreted as the redshift distribution of sources that, if lensed by CMASS galaxies, are sufficiently bright to be detected in a strong lensing survey like ours.
It is, therefore, a term related to the strong lensing selection function. As such, we could in principle express it as the product of two terms: one describing the redshift distribution of all possible source galaxies, multiplied by a sensitivity function, to be included in the term $\fsel$, which cuts off objects that are too faint to be detected. In practice, we choose to only model the product of these two terms, for the sake of convenience. By doing so, we are assuming that the selection in source redshift due to the sensitivity limit of our survey can be separated in a term that is independent of the lens model parameters (which is implicitly included in $\mathcal{Z}_s$) and a term that depends on the properties of the lens galaxy.
This latter term is not included in the simplest version of our model, but will be introduced in \Sref{sect:deteff}.
The term $\mathcal{Z}_s$, too, is modelled as a Gaussian distribution in $z_s$:
\begin{equation}
\mathcal{Z}_s(z_s|\hyperp) = \frac{1}{\sqrt{2\pi}\sigma_s}\exp{\left\{-\frac{(z_s - \mu_s)^2}{2\sigma_s^2}\right\}}.
\end{equation}

In summary, we model the distribution of individual object parameters of CMASS galaxies as the product of eight terms, listed in \Eref{eq:cmassterms}. The values of some of these hyper-parameters, those that have been measured with high precision, are kept fixed for the sake of reducing the dimensionality of the problem.
The list of free hyper-parameters then reduces to the set
\begin{equation}
\hyperp \equiv \{\mu_{h,0}, \sigma_h, \beta_h, \mu_{\mathrm{IMF}}, \sigma_{\mathrm{IMF}}, \mu_s, \sigma_s\}.
\end{equation}
We refer to this as the `base model'.

\subsection{Strong lensing selection}\label{ssec:selfunc}

The $\Nlens$ galaxies in our sample are not randomly selected from the general distribution of CMASS galaxies: they are strong lenses. The probability of a galaxy being a strong lens increases with increasing mass and, at fixed mass, with increasing concentration.
The term $\fsel$ then, which re-weights the distribution of CMASS galaxies by the probability of a galaxy-source pair to be included in a strong lens survey, must be proportional to $\crosssect$:
\begin{equation}
\fsel \propto \crosssect.
\end{equation}
The strong lensing cross-section of a lens-source pair is proportional to the area of the source plane that gets mapped into multiple images that are detectable by a strong lens survey.

To calculate $\crosssect$ in the context of our model we make a series of simplifying assumptions. 
We first assume circular symmetry. 
Secondly, following \citet{Son++18b}, we define $\crosssect$ as the angular size of the region of the source plane that gets mapped into sets of at least two images with magnification larger than a minimum value $|\mu_{min}|=0.5$. We verified that changing the value of $\mu_{min}$ does not change our results significantly.
Finally, we ignore any additional selection effect due to the lens-finding efficiency of the strong lens surveys on which our sample is based.
We will relax this assumption in \Sref{sect:deteff}.

\subsection{Inferring the hyper-parameters}

We wish to infer the posterior probability distribution of the hyper-parameters $\hyperp$ given the data. Using Bayes' theorem, this is
\begin{equation}
\pr(\hyperp|\data) \propto \pr(\hyperp)\pr(\data|\hyperp),
\end{equation}
where $\pr(\hyperp)$ is the prior probability distribution of the hyper-parameters, to be assigned, and $\pr(\data|\hyperp)$ is the likelihood of observing the data given the hyper-parameters.
Since the measurements on individual galaxies are all independent from each other, the latter can be expanded as
\begin{equation}
\pr(\data|\hyperp) = \prod_i \pr(\datai|\hyperp),
\end{equation}
where $\datai=\{\teini,\mchabi\}$ is the measurement of the Einstein radius and stellar mass of the $i-$th lens, inclusive of their uncertainties.
With $\individi$ being the set of individual parameters of the $i-$th lens, listed in \Eref{eq:individ},
each term of the product in the right-hand side of the above equation can then be written as
\begin{equation}\label{eq:integral}
\pr(\datai|\hyperp) = \int d\individi \pr(\datai|\individi) \pr(\individi|\hyperp).
\end{equation}
In other words, the likelihood of the data given the hyper-parameters is obtained by marginalising over all possible values of the individual parameters of the $i-$th lens, with a weight $\pr(\individi|\hyperp)$ given by the hyper-parameters, which effectively acts as a prior on $\individi$. This last term is the product of \Eref{eq:psplit} between the distribution of individual object parameters of CMASS galaxies and the strong lensing selection term, which, as discussed in the previous subsection, is proportional to the strong lensing cross-section:
\begin{equation}\label{eq:hyperdist}
\pr(\individi|\hyperp) = A(\hyperp)\crosssect(\individi) \pcmass(\individi|\hyperp).
\end{equation}
The factor $A(\hyperp)$ is a multiplicative constant ensuring that $\pr(\individ|\hyperp)$ is normalised to unity.

We sample the posterior probability distribution with an MCMC, using {\sc emcee}.
At each step of the chain, we calculate the multi-dimensional integral of \Eref{eq:integral} with importance sampling and Monte Carlo integration.
The full procedure is described in Appendix~\ref{sect:appendixa}.

\subsection{The prior}

We assume flat priors on $\mu_{\mathrm{IMF}}$ and $\sigma_{\mathrm{IMF}}$, describing the distribution in the IMF mismatch parameter.
The remaining free hyper-parameters are $\mu_{h,0}$, $\sigma_h$ , and $\beta_h$, describing the distribution in halo mass. For these, we set a prior based on the weak lensing study of S19.

The S19 measurement was obtained assuming a Chabrier IMF for all CMASS galaxies, while here we let the IMF normalisation to be a free parameter. We expect the inference of the halo mass distribution from weak lensing to be mostly insensitive to the particular choice of the IMF, since this only affects the mass in the very inner regions of each lens, while the weak lensing data used for their analysis extends out to $300$~kpc.
To verify this conjecture, we repeat the S19 analysis assuming different, but fixed, values for $\aimf$ of CMASS galaxies.
A $0.1$~dex increase in $\aimf$ results in a $0.03$ decrease in the maximum-likelihood value of $\mu_{h,0}$, while the inference on $\sigma_h$ and $\beta_h$ is unchanged.

Given the small effect of a varying IMF on the weak lensing-based inference on the halo mass distribution, as a prior on $\{\mu_{h,0}, \sigma_h, \beta_h\}$ we use the posterior probability distribution inferred from weak lensing by assuming an IMF normalisation $\log{\aimf}=0.1$.
We then iteratively repeat the weak lensing measurement by setting $\aimf$ to the maximum-likelihood value of $\mu_{\mathrm{IMF}}$ obtained in our inference, until the result is stable (in practice, the procedure converges at the first iteration).
For consistency with the model used in this work, we also assume that the average halo mass depends only on stellar mass, as specified by \Eref{eq:muhalo}, and not on the half-light radius as was assumed by S19. This is justified by the fact the S19 did not find any evidence for an additional dependence of halo mass on size at fixed stellar mass.
Under these assumptions, the new inference on the three hyper-parameters describing the halo mass distribution is $\mu_{h,0} = 12.75\pm0.03$, $\sigma_h=0.33\pm0.03,$ and $\beta_h = 1.82\pm0.11$.
We approximate the posterior probability distribution of this weak lensing-based inference as a tri-variate Gaussian, with covariance matrix set equal to the covariance of the samples of the MCMC in the three parameters, and use it as a prior for our strong lensing inference.

\subsection{Results}\label{ssec:results}

\begin{figure*}
\includegraphics[width=\textwidth]{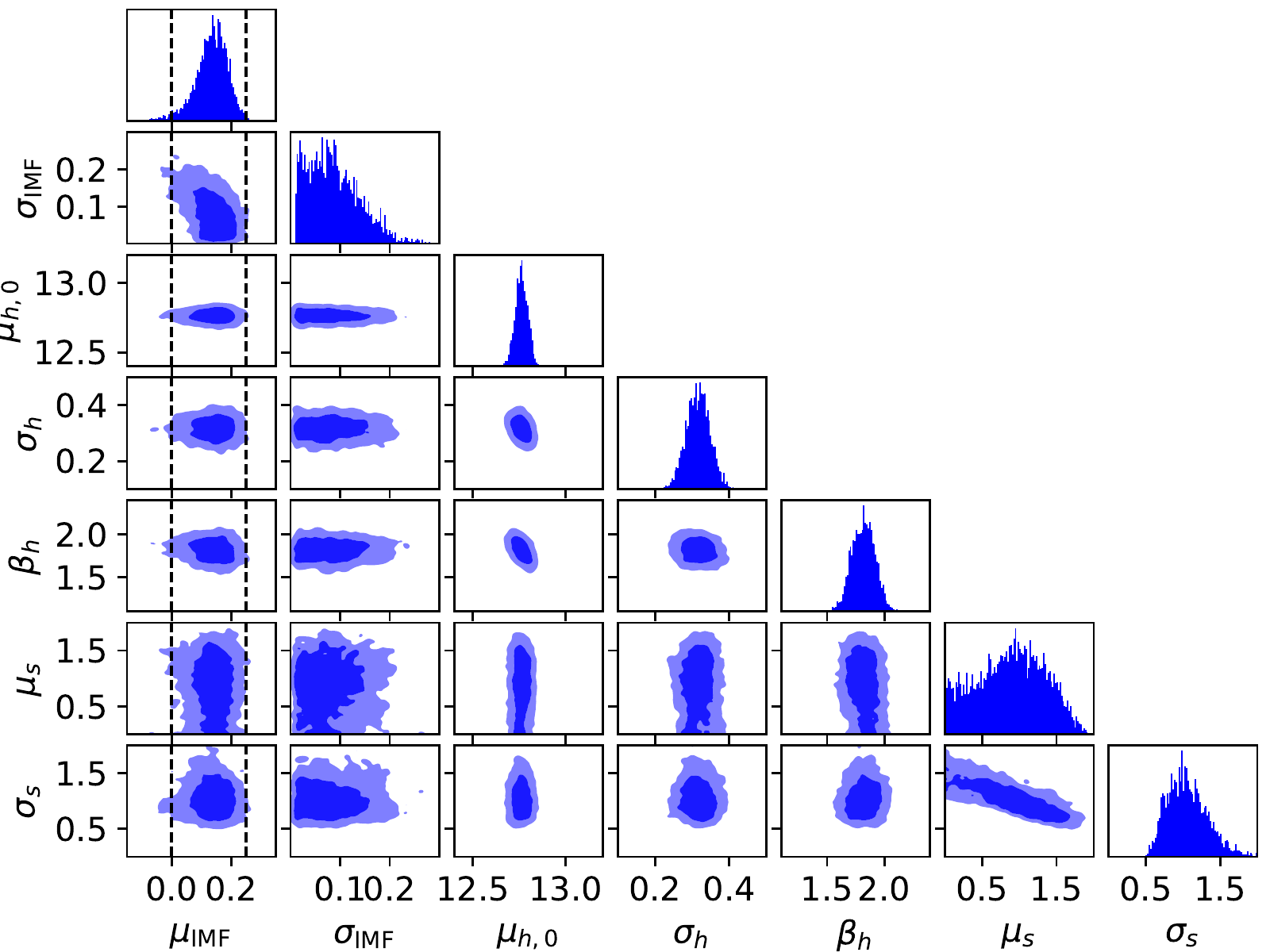}
\caption{Posterior probability distribution of the model hyper-parameters.
Contour levels correspond to the 68\% and 95\% enclosed probability regions.
The two vertical dashed lines on the first column, parameter $\mu_{\mathrm{IMF}}$, indicate an average IMF normalisation corresponding to a Chabrier and a Salpeter IMF, respectively.
\label{fig:nfw}
}
\end{figure*}

In \Fref{fig:nfw}, we show the posterior probability distribution on the model hyper-parameters. 
The median and 68\% enclosed marginal probabilities of each hyper-parameter are listed in the first column of \Tref{tab:inference}.

\begin{table*}
\caption{Inferred values of the hyper-parameters, with $68\%$ credible intervals. The first column refers to the inference obtained with the base model, introduced in \Sref{sect:method}. Values in the second and third column refer to the models with detection efficiency correction introduced in \Sref{sect:deteff}.}
\label{tab:inference}
\begin{tabular}{lcccl}
\hline
\hline
Parameter & Base model & `Arctan` model & `Gaussian` model & Description \\
\hline
$\mu_{h,0}$ & $12.76 \pm 0.03$ & $12.77 \pm 0.03$ & $12.76 \pm 0.03$ & Mean $\log{\mhalo}$ at $\log{\mchab}=11.4$ \\
$\sigma_h$ & $0.32 \pm 0.03$ & $0.32 \pm 0.03$ & $0.32 \pm 0.03$ & Scatter in $\log{\mhalo}$ around the mean \\
$\beta_h$ & $1.79 \pm 0.10$ & $1.78 \pm 0.10$ & $1.82 \pm 0.11$ & Power-law dependence of halo mass on $\mchab$ \\
$\mu_{\mathrm{IMF}}$ & $0.14 \pm 0.05$ & $-0.04 \pm 0.11$ & $-0.03 \pm 0.12$ & Mean $\log{\aimf}$ \\
$\sigma_{\mathrm{IMF}}$ & $0.06 \pm 0.05$ & $0.10 \pm 0.07$ & $0.11 \pm 0.07$ & Scatter in $\log{\aimf}$ \\
$\mu_s$ & $0.89 \pm 0.51$ & $0.58 \pm 0.41$ & $0.56 \pm 0.39$ & Mean source redshift \\
$\sigma_s$ & $1.06 \pm 0.29$ & $0.96 \pm 0.16$ & $0.96 \pm 0.18$ & Scatter in the source redshift \\
$\log{m}$ & $\ldots$ & $1.55 \pm 0.56$ & $\ldots$ & Lens detection efficiency parameter (`arctan` model) \\
$\theta_0$ & $\ldots$ & $0.95 \pm 0.09$ & $\ldots$ & Lens detection efficiency parameter (`arctan` model) \\
$\mu_{\theta}$ & $\ldots$ & $\ldots$ & $1.49 \pm 0.14$ & Lens detection efficiency parameter (`Gaussian` model) \\
$\sigma_\theta$ & $\ldots$ & $\ldots$ & $0.32 \pm 0.10$ & Lens detection efficiency parameter (`Gaussian` model) \\

\end{tabular}
\end{table*}

Our data, combined with the weak lensing prior, result in an inferred value of the average $\log{\aimf}$ of $\mu_{\mathrm{IMF}} = 0.14\pm0.05$. This value is in between that of a Chabrier IMF (corresponding to $\log{\aimf}=0$ by definition) and a Salpeter IMF (corresponding to $\log{\aimf}\approx0.25$, for quiescent galaxies like the ones in our sample).

\subsection{Goodness-of-fit evaluation}\label{ssec:pptest}

In Bayesian hierarchical inference studies, goodness-of-fit is evaluated by means of posterior predictive tests: we use the posterior probability distribution of the model to generate mock data, then compare this generated data with observations on the basis of test quantities summarising the mismatch between the datasets.
In our case, the key observable is the sample of Einstein radii, $\{\theta_{\mathrm{Ein},i}\}$.
We then generate sets of $\Nlens$ lens-source pairs, calculate their Einstein radii, and compare them to the distribution of the observed values.

We consider four test quantities $T_1,\ldots,T_4$: $T_1$ and $T_2$ are the mean and standard deviation of the Einstein radius distribution, while $T_3$ and $T_4$ are the minimum and maximum value of the Einstein radius in the sample.
The observed values of these test quantities are $T_1^{\mathrm{(obs)}} = 1.283''$, $T_2^{\mathrm{(obs)}} = 0.319''$, $T_3^{\mathrm{(obs)}} = 0.724'',$ and $T_4^{\mathrm{(obs)}} = 2.143''$.
For each test quantity $T_n$, we wish to determine the probability of the model predicting a more extreme value than the observed one:
\begin{equation}
\pr(T_n^{\mathrm{(pp)}} \lessgtr T_n^{\mathrm{(obs)}}),
\end{equation}
where $T_n^{\mathrm{(pp)}}$ stands for the posterior predicted test quantity, and the probability must be calculated by averaging over the posterior probability distribution.

We proceed as follows: we randomly draw 1,000 points from the MCMC sample of the posterior probability distribution, then, for each point, generate a large sample of galaxy-source pairs from the model distribution $\pcmass$. We then assign a probability proportional to $\fsel$ to each galaxy-source pair in this sample and draw $\Nlens$ lenses. For each set of mock lenses, we calculate the posterior predictive test quantities. 

\begin{figure*}
\includegraphics[width=\textwidth]{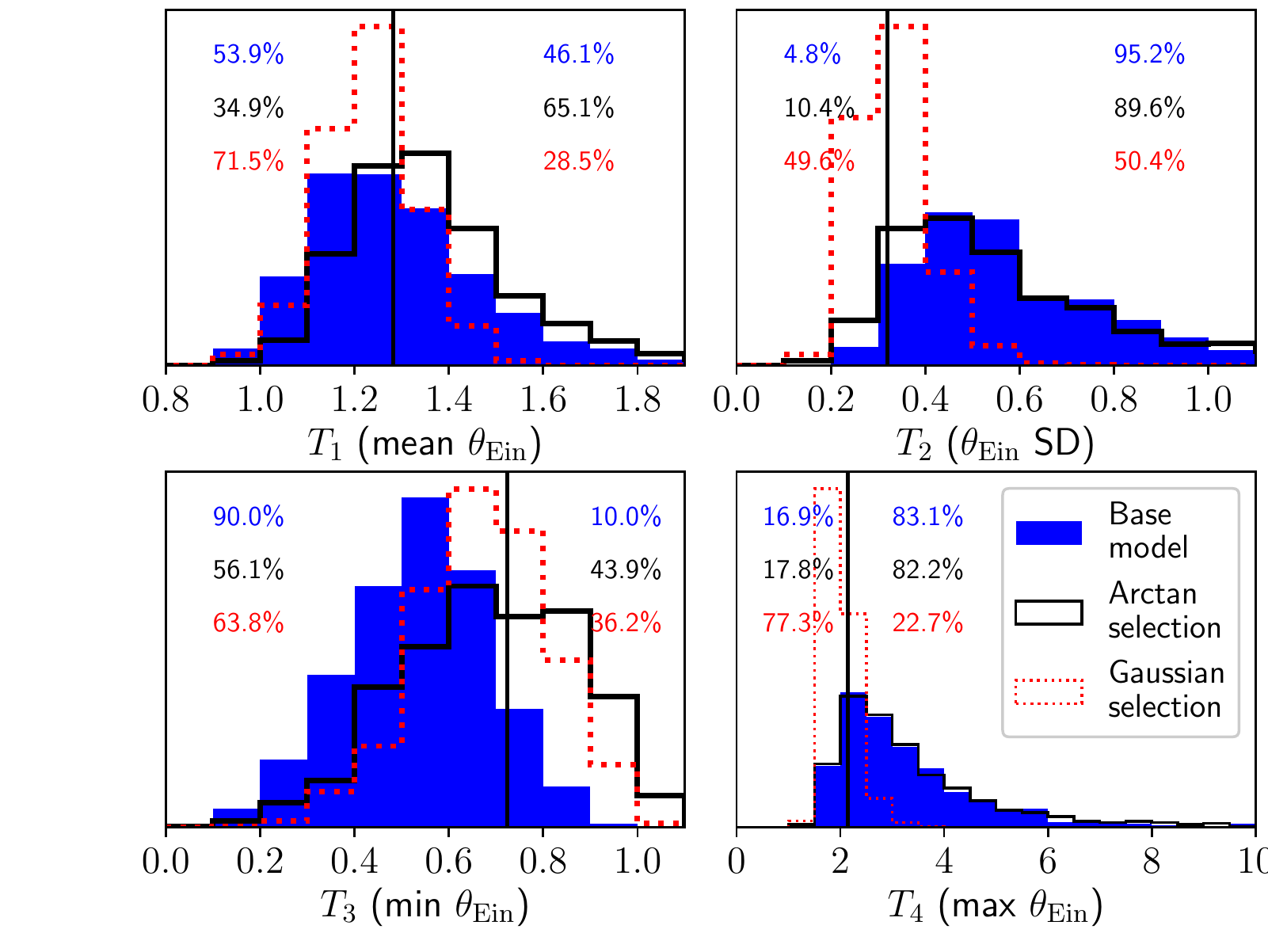}
\caption{Posterior predictive tests. The four panels show predicted distributions in the average, standard deviation, minimum, and maximum $\tein$ (in arcseconds) in samples of $\Nlens$ strong lens systems drawn from the posterior. The vertical line in each panel corresponds to the observed value of each test quantity. 
Posterior predicted distributions obtained from the base model are shown as filled histograms in blue. The black solid line and red dotted line histograms correspond to the `Arctan` and `Gaussian` lens detection efficiency models introduced in \Sref{sect:deteff}. The percentage to the left (right) side of each panel indicates the fraction of times the predicted value is smaller (larger) than the observed one,
with the top value corresponding to the base model and the middle and bottom values corresponding to the `Arctan` and `Gaussian` models, respectively.
\label{fig:pptest}
}
\end{figure*}

In \Fref{fig:pptest}, we plot the posterior predictive distribution of the four test quantities. 
Under the assumption that our model is an accurate description of reality, the value of the observed average Einstein radius is typical, with the posterior predicted values exceeding it 46.1\% of the time. The predicted values of the standard deviation in the Einstein radius, though, are mostly larger than the observed value: if the model is correct, the probability of observing a value of the standard deviation equal to $0.319''$ or smaller is only 4.8\%. This low probability casts a doubt on the ability of the model to accurately describe this aspect of the data, although there is a non-negligible possibility that our sample of lenses has an unusually small value of the standard deviation in $\tein$ by pure chance.

The posterior prediction on the minimum value of the Einstein radius of the sample ($T_3$, bottom left panel of \Fref{fig:pptest}) shows that the model typically predicts smaller values than $T_3^{\mathrm{obs}}$: the observed value of $0.724''$ is exceeded in only 10.0\% of the posterior draws.
Finally, the test on the maximum value of the Einstein radius, $T_4$, is less conclusive, with the model predicting smaller values than the observed one 16.9\% of the time.

\section{Inferring the lens detection efficiency}\label{sect:deteff}

The posterior predictive tests carried out in Sect. \ref{ssec:pptest} suggest that the model, when used to generate mock observations of samples of $\Nlens$ strong lenses, tends to predict a broader distribution in $\tein$ than observed, with more than 95\% significance.
Additionally, the minimum value of $\tein$ of the observed sample, $0.724''$, is relatively large when compared to the posterior predicted distribution of the same quantity.

We wish to add complexity to our model in order to alleviate these mild tensions between posterior predicted quantities and observations.
One important aspect of the problem that has not yet been taken into account is observational selection effects related to differences in the detection efficiency of lenses with different properties. For example, strong lenses with a small Einstein radius are more difficult to identify in ground-based imaging data, where most of the systems used for this study have been discovered, due to the effects of atmospheric blurring.
The observed narrow distribution in $\tein$ and the relatively large value of the minimum Einstein radius of the sample could be the result of this observational effect.
We then modify the strong lensing selection term, $\fsel$, by multiplying the lensing cross-section by a detection efficiency term,
\begin{equation}
\fsel = \crosssect \fdet(\individ),
\end{equation}
which, in general, can be a function of all the model parameters. 
Strictly speaking, it is not appropriate to define a single detection efficiency for the whole sample of lenses. This is because not all lenses have been selected homogeneously from a single survey. Some of them have been discovered in data from the Canada-France-Hawaii Telescope \citep{Gav++14}, some from the HSC survey, while for some others, those belonging to the BELLS survey, HST data have been used for their confirmation.
Moreover, the properties of the sample depend also on the spectroscopic data used to measure the redshift of the background source. For a good fraction of our lenses, the source redshift is obtained directly from the BOSS spectrum: since the BOSS fibre has a $1''$ radius, lenses with a much larger value of $\tein$ are less likely to have their source detected \citep{Arn++12}.
Finally, we have required in Sect. \ref{ssec:sample} that all the lenses consist of a single galaxy as the deflector. This condition tends to exclude lenses with a large Einstein radius, because these are more likely to have close multiplets of galaxies acting as a lens.

It is prohibitively difficult to write down an analytical form for $\fdet(\individ)$ that takes into account all of these different selection effects.
Instead, we make the simplifying assumption that the detection efficiency of the whole lens sample can be described by an analytical function of the Einstein radius:
\begin{equation}
\fdet(\individ) \approx \fdet(\theta).
\end{equation}
An implicit assumption in this approach is that, at fixed $\theta$, any additional dependence of the detection efficiency on the source redshift $z_s$ is fully captured by the term $\mathcal{Z}_s$ in $\pcmass$.
We explore two different choices for the form of $\fdet$, as described below:
\begin{equation}
\fdet(\theta) \propto \left\{\begin{array}{ll} \dfrac{1}{\pi} \arctan{\left[m(\tein - \theta_0)\right]} + \dfrac12 & \mathrm{`Arctan`\,model} \\
\exp{\left\{-\dfrac{(\tein - \mu_\theta)^2}{2\sigma_\theta^2}\right\}} & \mathrm{`Gaussian`\,model}
\end{array}\right.
.\end{equation} 
The `Arctan` detection efficiency model goes to zero for values of $\tein$ smaller than $\theta_0$ and reaches a constant for large $\tein$. The steepness of the transition between the two regimes depends on the parameter $m$.
The `Gaussian` model goes to zero both for small and large values of the Einstein radius.
The rationale for the low-$\tein$ cutoff present in both models is to capture the loss of lenses due to the resolution limit of HSC data.
The `Gaussian` model has an additional cutoff at large values of $\tein$, which is meant to describe a loss due to our selection of isolated galaxies as lenses, and the decrease in the success rate of source redshift measurements from BOSS spectroscopy at large $\tein$. 
Each of these models is described by two parameters, which we infer from the data.
We assume a flat prior on $\theta_0$, $\log{m}$, $\mu_\theta,$ and $\sigma_\theta$.
With the addition of the multiplicative term $\fdet$, the probability distribution $\fsel\pcmass = \crosssect\fdet\pcmass$ now describes the distribution of CMASS strong lenses that can be detected in a survey like ours. From here on, we will refer to this distribution as that of SuGOHI lenses.

\begin{figure*}
\includegraphics[width=\textwidth]{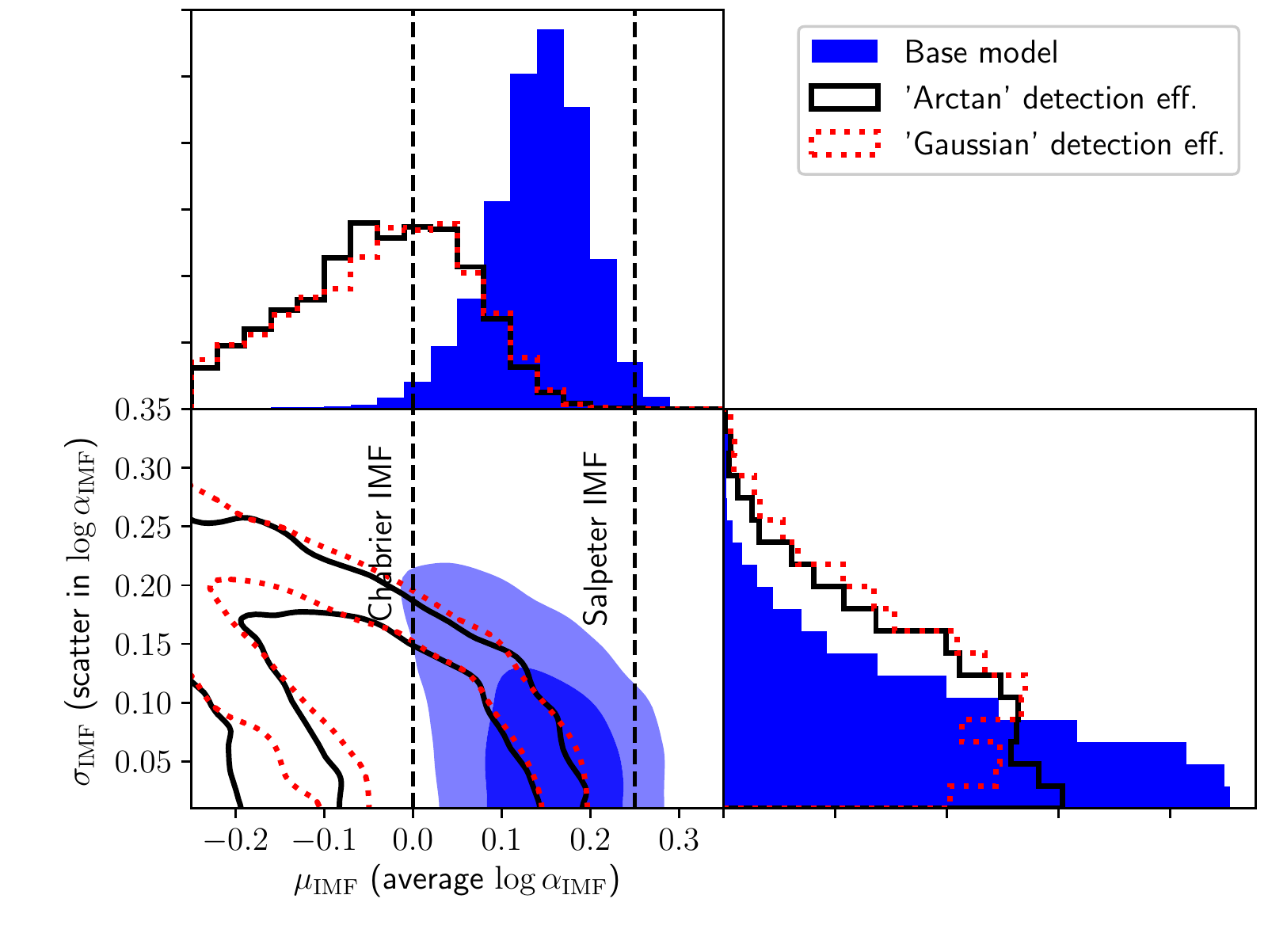}
\caption{Posterior probability distribution of the hyper-parameters describing the IMF normalisation of the CMASS sample, marginalised over the other hyper-parameters. {\em Filled contours}: Base model. {\em Black solid lines}: Model with `Arctan` detection efficiency function. {\em Red dotted lines}: Model with `Gaussian` detection efficiency function. Contour levels mark the 68\% and 95\% enclosed probability regions. The vertical dashed lines correspond to an average IMF normalisation equal to that of a Chabrier and Salpeter IMF, respectively.
\label{fig:selfuncimf}
}
\end{figure*}

In \Fref{fig:selfuncimf} we show the new inference on the pair of hyper-parameters describing the mean and intrinsic scatter in $\log{\aimf}$, obtained with the two different prescriptions for the lens detection efficiency. For comparison, we also plot the inference obtained in the previous section, with no detection efficiency correction (filled contours). The median values and 68\% credible regions of the full set of hyper-parameters, including those describing the detection efficiency, are reported in the second and third column of \Tref{tab:inference}.

The main effects of adding a lens detection efficiency correction $\fdet$ are to broaden and to shift towards smaller values the distribution of allowed values of the average IMF normalisation parameter $\mu_{\mathrm{IMF}}$. Remarkably, the inferences obtained with the two different forms for $\fdet$ are very similar to each other. This is reassuring as it indicates that the result is not particularly sensitive to the specific choice of functional form for $\fdet$.
For the `Arctan` model, the marginal posterior distribution in $\mu_{\mathrm{IMF}}$ has a median and 68\% confidence interval of $-0.04\pm0.11$, while the 95th percentile of the distribution is $0.13$. Numbers relative to the `Gaussian` model are very similar.
These result are inconsistent with an average IMF normalisation equal to that of a Salpeter IMF.

We now assess the goodness-of-fit of these two new models, with the same method used in Sect. \ref{ssec:pptest}.
The posterior predictive distributions of the four test quantities $T_1,\ldots,T_4$ obtained from the `Arctan` and `Gaussian` model are plotted in \Fref{fig:pptest}.
For both of these models, in none of the test quantities the observed value is more extreme than the lowest or highest 10\% tail of the posterior predicted distribution. This indicates that 1) both models are able to reproduce all four aspects of the observed distribution in $\tein$ and 2) the data do not give us strong reasons to favour one model of the lens detection efficiency over the other.

\section{Discussion}\label{sect:discuss}

The statistical combination of strong lensing measurements on a sample of $\Nlens$ CMASS galaxies and weak lensing constraints on the distribution of halo mass as a function of stellar mass of the parent sample, allows us to infer the average IMF mismatch parameter of CMASS galaxies.
We find a value consistent with that of a Chabrier IMF, while a normalisation as heavy as that of a Salpeter IMF is in clear tension with our inference.
In the following subsections we discuss how sensitive these results are to the various assumptions made in our analysis, how they compare with similar studies from the literature, and implications for the relative distribution of strong lenses and non-lenses, given our model.

\subsection{Sensitivity to model assumptions}\label{ssec:assumpt}

One of the key assumptions in our model is that of an NFW density profile for the dark matter halos in the sample, which allows us to use weak lensing information, obtained on scales larger than $\sim50$~kpc, to predict the projected dark matter mass enclosed within the Einstein radius of the lenses in our sample and thus separate the contribution of luminous and dark matter to the observed lensing masses.
If we were to lift this assumption, the inferred IMF mismatch parameter would be highly degenerate with the inner dark matter distribution.
In particular, for the same total mass, a halo with an overall steeper density profile, such as an adiabatically contracted one, corresponds to a higher central density of dark matter, implying that a lower stellar mass (and therefore a lower IMF normalisation) would be required to reproduce the observed value of the Einstein radius of a lens \citep[see  also][]{Aug++10}. This would increase the tension between our measurement and a scenario with an IMF normalisation equal to that of a Salpeter IMF.
Vice versa, this tension would decrease by allowing for the inner slope of the dark matter halo to be shallower than that of an NFW profile.

There is some observational evidence suggesting that the inner dark matter density profile of massive galaxies is generally steeper than the NFW model predicted by dark-matter-only simulations \citep{Son++12, O+A18b}. A similar behaviour is seen in the latest cosmological hydrodynamical simulations \citep{Xu++17, Pei++17}.
There are, however, also measurements indicating inner dark matter profiles flatter than NFW for some massive ETGs \citep{Bar++13, O+A18b}. The issue of whether dark matter halos of massive ETGs are more or less concentrated than standard dark-matter-only NFW halos of the same mass is therefore still subject to debate.

Our results depend also, in principle, on the assumed functional form of the distribution of structural parameters across the population of CMASS galaxies, \Eref{eq:cmassterms}.
In particular, the dark matter halo mass distribution term, $\mathcal{H}(\mhalo)$, plays an important role in our analysis, because the dark matter mass is not directly constrained by the strong lensing data and we rely on our prior knowledge of $\mathcal{H}$, that is, the SHMR of CMASS galaxies inferred from weak lensing, to disentangle the luminous and dark matter contribution to the strong lensing mass.
In our model, $\mathcal{H}(\mhalo)$ is a power-law relation between halo mass and stellar mass, with scatter.
This is an approximation: the SHMR is typically described by a more flexible model, with a change in slope around stellar mass $\sim10^{11}M_\odot$ \citep[see e.g.][]{Lea++12}.
However, in the stellar mass range covered by CMASS galaxies, deviations from a pure power-law SHMR are smaller than the width of our prior (and the posterior) on $\mathcal{H}$ (i.e. varying the slope of the SHMR, parameter $\beta_h$, within its uncertainty has a bigger impact on the SHMR).
We then conclude that our results are not sensitive to our particular choice for the parameterisation of the SHMR.

Another assumption in our model is that of a spatially constant stellar IMF, which is in contrast with some recent spatially resolved studies of the IMF in massive galaxies (see \Sref{sect:intro}).
In order to assess the impact of this assumption on our inference, it is useful to generalise the definition of the IMF mismatch parameter $\aimf$ by introducing the enclosed IMF mismatch parameter profile, $\aimf(<R)$, defined as the ratio between the true stellar mass enclosed within radius $R$ and the stellar mass measured in the same region from stellar population synthesis modelling, assuming a Chabrier IMF \citep[see also Sect. 5 of][]{Son++18b}.
Strong lensing data is only sensitive to $\aimf(<R_{\mathrm{Ein}})$, where $R_{\mathrm{Ein}}$ is the Einstein radius in physical units.
If the true IMF mismatch parameter of a galaxy is declining with radius, then $\aimf(<R)$ is also a decreasing function of $R$, meaning that, on average, $\aimf(<R_{\mathrm{Ein}})$ is larger for lenses with a smaller Einstein radius and vice-versa.
In our population model we do not allow for such a trend. As a result, this signal, if present, would translate into a larger intrinsic scatter in the $\aimf$ distribution.
However, our inference on the IMF mismatch parameter intrinsic scatter $\sigma_{\mathrm{IMF}}$ is consistent with zero, meaning that if a signal from a spatially varying IMF is present in our sample, our measurements are not sensitive to it.

\subsection{Comparison with other IMF studies}\label{ssec:compare}

Under the assumption of an NFW dark matter density profile, we find an IMF mismatch parameter consistent with that of a Chabrier IMF, while a normalisation as heavy as that of a Salpeter IMF is excluded at more than $2\sigma$ level.
This value appears to be in tension with some measurements of the stellar IMF in strong lenses from the literature. For instance, \citet{Tre++10}, combining strong lensing with stellar dynamics on a set of 56 lenses from the Sloan Lenses ACS Survey (SLACS), claimed an average IMF normalisation higher than that of a Salpeter IMF. A similar result was found by \citet{Son++15}, using a similar method on a sample of 80 lenses drawn from the SLACS and the SL2S survey, and by \citet{Cap++12} from a spatially resolved dynamical study of a sample of nearby ETGs.

The recent re-analysis of the SLACS sample by \citet{Son++18b}, however, showed how stellar dynamics is particularly sensitive to the presence of gradients in the $M_*/L$ in the inner regions of galaxies. In particular, negative gradients in $M_*/L$ can bias the inferred IMF normalisation towards larger values than the truth: this is because, at fixed mass enclosed within the Einstein radius, a negative gradient in $M_*/L$ increases the central velocity dispersion, in a similar way as increasing the global IMF mismatch parameter does \citep[see also][]{Ber++18}.

When allowing for $M_*/L$ gradients and including weak lensing information, \citet{Son++18b} inferred an average IMF normalisation $\mu_{\mathrm{IMF}}=0.11\pm0.04$ for the SLACS lenses, which translates into even lower values when accounting for lensing selection effects.
Our inference is then consistent with that of \citet{Son++18b}, and in general with a series of recent studies based on spatially-resolved kinematics and strong lensing \citep{O+A18a, O+A18b, Col++18}, which show that the region in massive galaxies where the IMF normalisation is significantly heavier than that of a Chabrier IMF is limited to $R\lesssim1\,\rm{kpc}$.

If the sensitivity of stellar dynamics to $M_*/L$ gradients is the main reason for the discrepancy between our results and previous estimates of the IMF normalisation from strong lensing and dynamics, and if $M_*/L$ gradients are also present in CMASS galaxies, we would expect our gradient-less model to under-predict the central velocity dispersion of our lenses, compared to the values measured by BOSS.
We can verify this conjecture with a posterior predictive test and by making additional assumptions on the dynamical state of our lenses.

\begin{figure}
\includegraphics[width=\columnwidth]{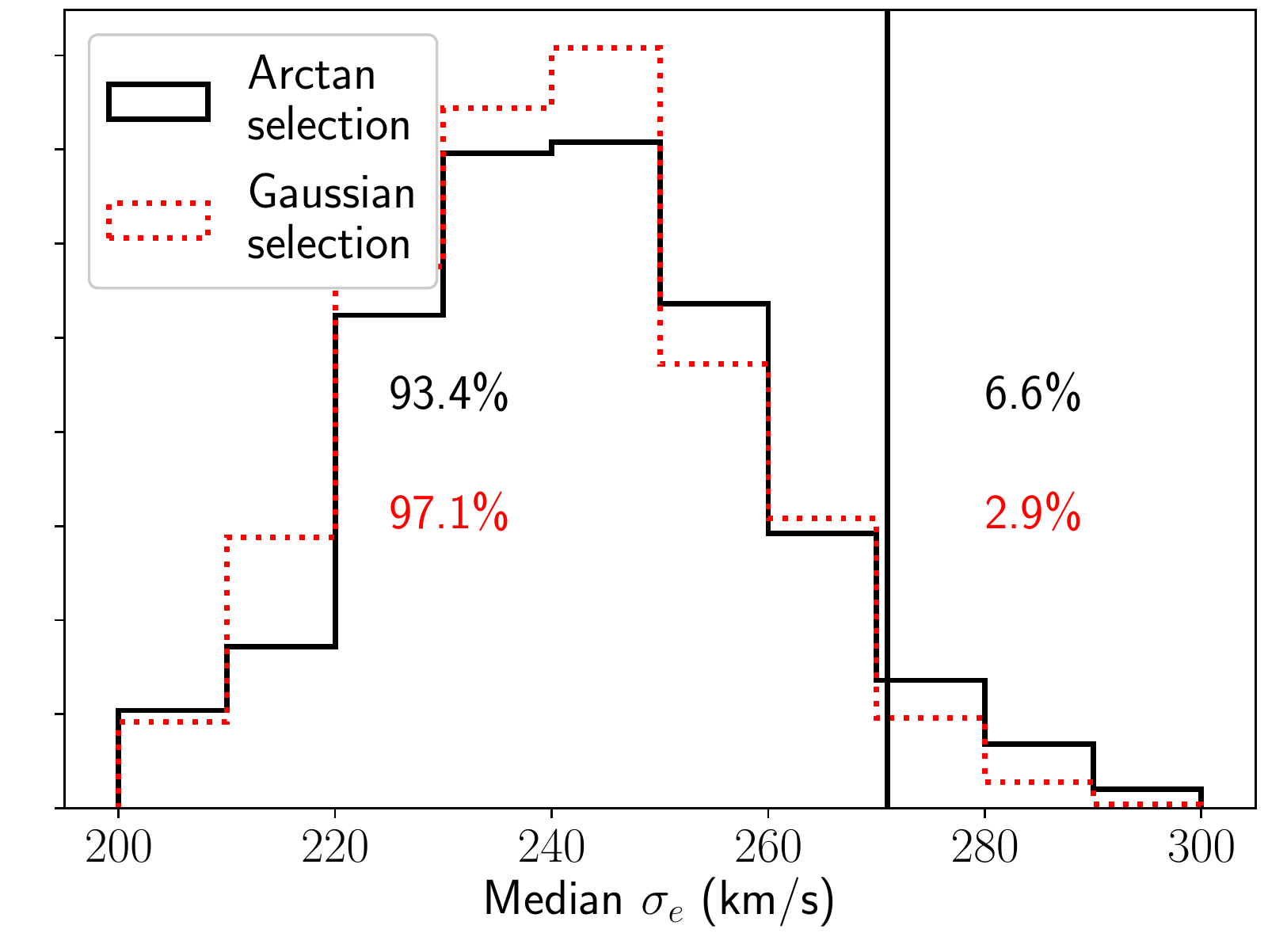}
\caption{Posterior predictive distribution in the median value of the central stellar velocity dispersion $\sigma_{\mathrm{e}}$, from mock realisations of sets of $\Nlens$ SuGOHI lenses, generated from the posterior probability distribution inferred for the `Arctan` (black solid) and `Gaussian (red dotted) models.
The vertical line marks the median value of the stellar velocity dispersion of the lenses in our sample, as measured by BOSS.
The percentages left and right of the vertical line indicate the fraction of mock realisations with a median $\sigma_{\mathrm{e}}$ smaller or larger than the observed value, for the two models.
\label{fig:sigmapp}
}
\end{figure}

We take the same mock realisations of sets of $\Nlens$ lenses used for the posterior predictive tests of \Fref{fig:pptest}, drawn from the posterior probability distribution of the `Arctan` and `Gaussian` models. Then, for each lens, we use the spherical Jeans equation, assuming isotropic orbits, to predict the surface brightness-weighted line-of-sight stellar velocity dispersion, integrated within a circular aperture of radius $\reff$: $\sigma_{\mathrm{e}}$\footnote{The code used for the calculation of the velocity dispersion through the spherical Jeans equation is available at the following link: \url{https://github.com/astrosonnen/spherical_jeans}.}.
For a fair comparison with the noisy distribution in observed velocity dispersion, we add random errors to the mock distribution in $\sigma_{\mathrm{e}}$, with a $46$~km~s$^{-1}$ scatter, which is the median value of the uncertainty on the BOSS velocity dispersion of our sample.
For each mock realisation, we then compare the median value of $\sigma_{\mathrm{e}}$ with the median $\sigma_{\mathrm{BOSS}}$ of our sample, which is $271$~km~s$^{-1}$.
We choose the median rather than the mean, because it is less sensitive to the presence of catastrophic outliers, such as the value of $\sigma_{\mathrm{BOSS}}$ for HSCJ121504$+$004726.
In \Fref{fig:sigmapp} we plot the posterior predicted distribution of the median $\sigma_{\mathrm{e}}$, based on 1,000 random realisations.
As expected, our model tends to under-predict the velocity dispersion of the sample, with a large ($\sim95\%$) probability.
Adding a negative gradient in $M_*/L$ would bring our model into better agreement with the observations.

\subsection{Differences between strong lenses and the general population}

An important part of our analysis is the separation of the probability distribution of the strong lenses into a term describing the distribution of the parent sample and a term proportional to the strong lensing cross-section and the lens detection efficiency, which allows us to correct for selection effects.
For a set of values of the hyper-parameters $\hyperp$, the term $\pcmass(\individi|\hyperp)$ describes the general population of CMASS galaxies. If multiplied by the strong lensing cross-section $\crosssect$, we obtain the distribution of CMASS lenses. Finally, multiplication by the lens detection efficiency $\fdet$ gives the population of SuGOHI lenses.
Each of these three populations will in general occupy different regions of parameter space.
We can use a posterior predictive procedure to investigate these possible differences.

\begin{figure*}
\includegraphics[width=\textwidth]{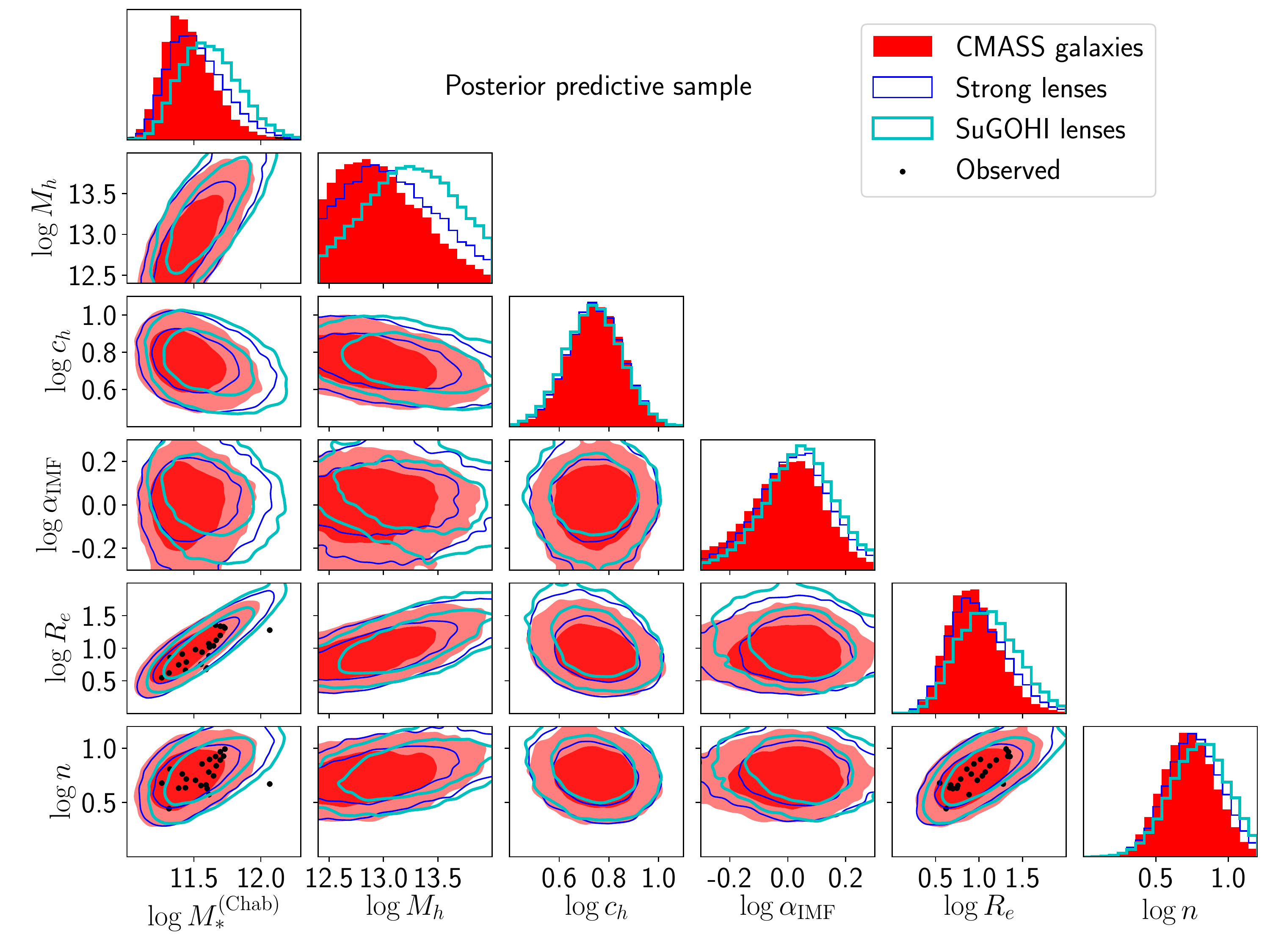}
\caption{Posterior predictive distribution in stellar mass, halo mass, halo concentration, IMF mismatch parameter, half-light radius, and S\'{e}rsic index of CMASS galaxies, CMASS strong lenses and SuGOHI lenses, obtained from our inference based on the `Arctan` model. Observed values in $\mchab$, $\reff,$ and $\nser$ of the $\Nlens$ lenses in our sample are also shown. Contour levels correspond to the 68\% and 95\% enclosed fraction.
\label{fig:ppcp}
}
\end{figure*}

We are interested in studying the posterior probability distribution of the individual galaxy parameters $\individ$, marginalised over all possible values of the hyper-parameters allowed by the data. For the distribution of CMASS galaxies, this is given by
\begin{equation}
\pcmass(\individ) = \int d\hyperp \pcmass(\individ|\hyperp) \pr(\hyperp|\data),
\end{equation}
while analogous distributions of the samples of strong lenses can be obtained by multiplying $\pcmass$ by the appropriate strong lensing selection terms ($\crosssect$ for CMASS lenses and $\crosssect\fdet$ for SuGOHI lenses).
We sample from $\pcmass(\individ)$ as follows: we first draw a set of values of the hyper-parameters $\hyperp$ from the posterior probability distribution, then draw a set of values of $\individ$ from $\pr(\individ|\hyperp)$, for a large number of iterations.
The resulting distribution for the CMASS galaxies, the CMASS strong lenses and the SuGOHI lenses, is shown \Fref{fig:ppcp}, as obtained from the `Arctan` detection efficiency model (the `Gaussian` model produces very similar results).

We focus on the following quantities: stellar mass, IMF mismatch parameter, halo mass and concentration, half-light radius, and S\'{e}rsic index.
The most obvious difference between the strong lens subsamples and the general population is that the former are on average more massive than the latter, in terms of both stellar and dark matter mass.
This is true for the population of strong lenses compared to all CMASS galaxies, but also for the SuGOHI lenses compared to the population of strong lenses, indicating that selection effects play an important role in the definition of our sample of lenses.

We can also see that the predicted stellar mass-size relation of the strong lenses appears to be steeper than that of the general population (see the panel in the first column from the left and second row from the bottom of \Fref{fig:ppcp}), with the lowest mass lenses being more compact compared to regular galaxies of the same mass.
This prediction appears to match the observed distribution of lenses in $\mchab-\reff$ space. A similar behaviour is seen, with higher statistical significance, in the SLACS sample of strong lenses when compared to the population of SDSS quiescent galaxies from which that sample is drawn \citep{SLACSX}.

The posterior predicted IMF mismatch parameter has a broad distribution, as a result of the large observational uncertainty (the posterior predicted distribution is obtained by marginalising over the whole posterior distribution for the model hyper-parameters), but we can still see how lenses tend to have higher values of $\aimf$ compared to the general population.
We point out, however, that models with no intrinsic scatter in $\aimf$ are allowed by the data: the probability distribution in $\sigma_{\mathrm{IMF}}$, shown in \Fref{fig:selfuncimf}, is consistent with the value $0$.
For those models, the IMF normalisation is the same for all galaxies, regardless of their strong lens nature.
Models with $\sigma_{\mathrm{IMF}} = 0$ are not represented in the posterior predicted distribution of \Fref{fig:ppcp} as a result of our choice of prior, which is uniform on values of $\sigma_{\mathrm{IMF}} > 0$.
If we were to impose a much more restrictive prior, for instance asserting a universal IMF (i.e. $\sigma_{\mathrm{IMF}} = 0$), then the difference in $\aimf$ between the general population of galaxies and the strong lenses would disappear by construction, while the model would still be providing a good fit to the data.

\begin{figure}
\includegraphics[width=\columnwidth]{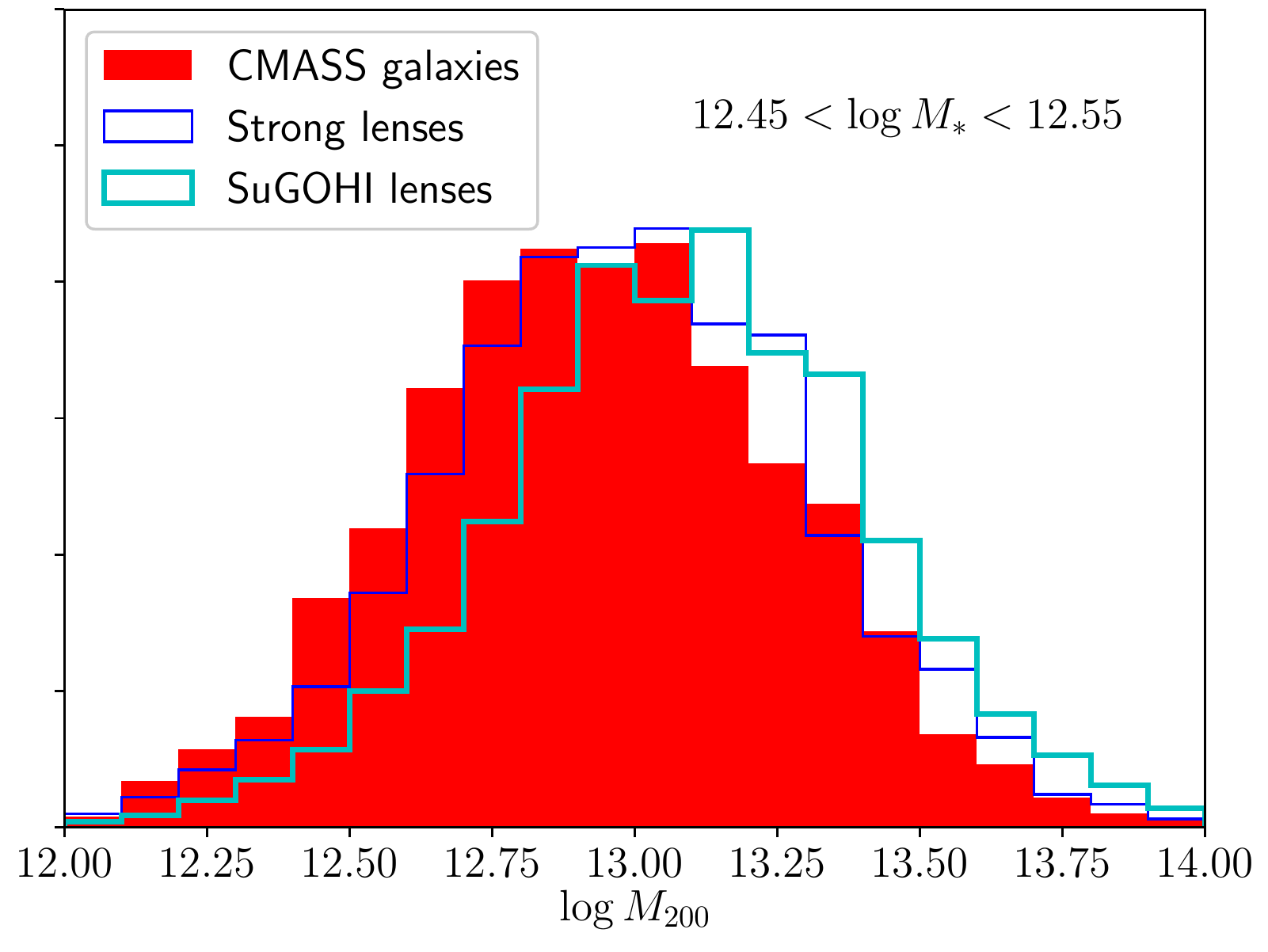}
\caption{Posterior predictive distribution in halo mass, for CMASS galaxies, CMASS strong lenses, and SuGOHI lenses with stellar masses in the range $11.45 < \log{\mchab} < 11.55$.
\label{fig:m200hist}
}
\end{figure}

Finally, the model predicts a higher average halo mass at fixed stellar mass for the strong lenses. This is best seen in \Fref{fig:m200hist}, where we plot histograms of the distribution in $\mhalo$ of posterior predicted galaxies and strong lenses selected in a narrow bin in stellar mass centred on $\log{\mchab} = 11.5$.
The halo mass of the strong lenses is on average $0.17$~dex higher than that of the general population of galaxies.
This model prediction can in principle be verified by carrying out a weak lensing analysis of CMASS strong lenses. In practice, this requires a much larger sample of lenses to reach the necessary precision to measure the signal.

\begin{figure}
\includegraphics[width=\columnwidth]{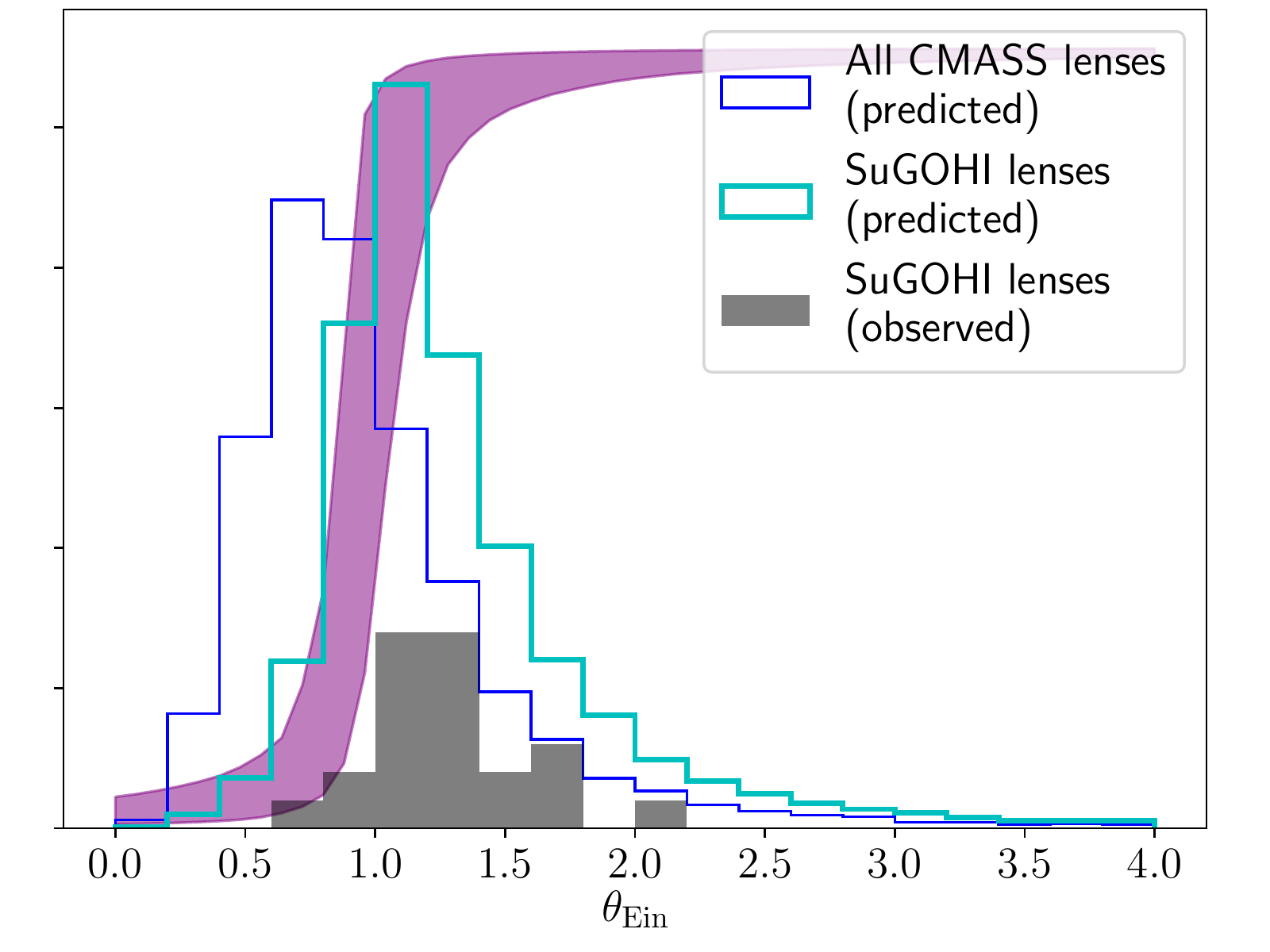}
\caption{Posterior predictive distribution in the Einstein radius of CMASS strong lenses and SuGOHI lenses, based on the `Arctan` model inference. The shaded curve corresponds to the 68\% credible region of the lens detection efficiency function, $\fdet$, as inferred from our data.
\label{fig:teinpp}
}
\end{figure}

We can also investigate the differences in the posterior predicted distribution in the Einstein radius between the SuGOHI lenses and all CMASS strong lenses.
This is shown in \Fref{fig:teinpp} for the `Arctan` model, together with the corresponding detection efficiency function $\fdet$, as inferred from the data.
Because of the sharp cutoff in $\fdet$ around $\tein\approx0.8''$, the distribution in $\tein$ of SuGOHI lenses is shifted towards larger values compared to that of all strong lenses: the medians of the two distributions are $1.18''$ and $0.88'',$ respectively. 
According to our model, then, half of the existing strong lenses in the CMASS sample have values of the Einstein radius smaller than $0.88''$, but many of these lenses are missed by our survey.

We can qualitatively test this prediction by considering the BELLS lens sample, which consists mostly of CMASS galaxy lenses and has minimal overlap with SuGOHI. The BELLS sample has been assembled by means of a spectroscopic search, and the lens detection efficiency of this survey extends to lower values of $\tein$ compared to ours \citep{Arn++12}. This implies that, if the population of small Einstein radius lenses predicted by our model exists, these lenses should be part of the BELLS sample. 
Indeed, half of the BELLS lenses have values of $\tein \leq 0.75''$, in agreement with our prediction. 
For a more quantitative comparison between our SuGOHI sample-based inference and the BELLS sample, it is necessary to model the lens detection efficiency of BELLS. This, however, is beyond the scope of this work.

\section{Conclusions}\label{sect:concl}

We used photometric data from HSC and spectroscopic measurements from VLT to analyse a sample of $\Nlens$ strong lenses drawn from the CMASS sample of BOSS galaxies.
We measured the Einstein radius and the stellar mass of each lens, then carried out a statistical analysis to infer the distribution in IMF normalisation of the CMASS sample of galaxies.
We used a prior on halo mass from a previous weak lensing analysis of the CMASS sample and assumed an NFW density profile for the dark matter distribution to break the degeneracy between the stellar IMF and the contribution of dark matter to the strong lensing mass. 
We also accounted for strong lensing selection effects to generalise the constraints obtained on the $\Nlens$ strong lenses to the parent sample.
In particular, our model accounts for the facts that 1) the probability of a galaxy being a lens is proportional to its strong lensing cross-section and 2) lenses with different Einstein radius have different probabilities of being detected by our lensing survey.

We constrain the average base-10 logarithm of the IMF normalisation of CMASS galaxies to be $-0.04\pm0.11$, where $\log{\aimf}=0$ corresponds to a Chabrier IMF, while a Salpeter IMF is in tension with our measurement.
This tension can be made more (less) severe by allowing the inner slope of the dark matter halo to be steeper (shallower) than that of an NFW profile.
Our inferred IMF normalisation is significantly lower than previous studies based on stellar dynamics, used either alone or in combination with strong lensing.
This discrepancy can be explained by the presence of radial gradients in the stellar mass-to-light ratio, to which stellar dynamics is particularly sensitive.
Ours is an estimate of the mass-weighted average IMF normalisation measured over a region of $5-10$~kpc in projection. As such, our measurement cannot rule out the presence of a stellar population with a heavier IMF confined to the very inner regions of the CMASS galaxies.

We investigated differences between the general population of CMASS galaxies, CMASS strong lenses, and lenses that can be detected by our survey, by means of posterior prediction.
Our model correctly predicts a steeper stellar mass-size relation for the strong lenses compared to the general population of galaxies, matching existing observations, and also predicts a higher average halo mass at fixed stellar mass for the lenses.

The current number of CMASS strong lenses (grade B or above) with HSC data, as of the 17A internal release, is 84, and is expected to at least double by the end of the HSC survey. Our sample size of $\Nlens$ was limited by the availability of spectroscopic measurements of the redshift of the background source.
With a larger sample of HSC lenses we expect to be able to relax some of our assumptions and test more complex models than the ones considered so far, such as models with a free inner density slope for the dark matter halo.

In the next decade, the Euclid space telescope\footnote{\url{https://euclid-ec.org/}} and the Large Synoptic Survey Telescope\footnote{\url{https://lsst.org}} (LSST) will enable the discovery of tens of thousands of new strong lenses \citep{Col15}. 
Together with a significant shrinkage in the statistical errors, such a dramatic increase in sample size will allow us to precisely infer the distribution of halo masses of the strong lenses directly from weak lensing, as opposed to using weak lensing information obtained on a separate sample as a prior for the strong lens sample as done in the present work.
This will then allow us to remove one source of potential systematic uncertainty, the strong lensing selection correction (although an accurate understanding of selection effects will still be needed to generalise results obtained on strong lenses to the general galaxy population).
Additionally, thanks to the high cadence of its planned observations, the LSST will provide time-delay measurements for hundreds of strongly lensed quasars \citep{O+M10}. Time-delay measurements are sensitive to higher order derivatives of the lens potential, compared to the image position data on which our work is based, and will then provide additional constraints on the density profile of the lens population. 
We expect the statistical strong lensing method developed in this work to be an essential tool to take full advantage of the wealth of strong lensing data coming from the next generation of surveys. 

\begin{acknowledgements}

AS acknowledges funding from the European Union's Horizon 2020 research and innovation programme under grant agreement No 792916, as well as a KAKENHI Grant from the Japan Society for the Promotion of Science (JSPS), MEXT, Number JP17K14250.
ATJ is supported by JSPS KAKENHI Grant number JP17H02868.
MO acknowledges support from JSPS KAKENHI Grants Number JP15H05892 and JP18K03693. SHS thanks the Max Planck Society for support through the Max Planck Research Group.
KCW is supported in part by an EACOA Fellowship awarded by the East
Asia Core Observatories Association, which consists of the Academia
Sinica Institute of Astronomy and Astrophysics, the National
Astronomical Observatory of Japan, the National Astronomical
Observatories of the Chinese Academy of Sciences, and the Korea
Astronomy and Space Science Institute.
This work was supported by World Premier International Research Center Initiative (WPI Initiative), MEXT, Japan. 

The Hyper Suprime-Cam (HSC) collaboration includes the astronomical communities of Japan and Taiwan, and Princeton University.  The HSC instrumentation and software were developed by the National Astronomical Observatory of Japan (NAOJ), the Kavli Institute for the Physics and Mathematics of the Universe (Kavli IPMU), the University of Tokyo, the High Energy Accelerator Research Organization (KEK), the Academia Sinica Institute for Astronomy and Astrophysics in Taiwan (ASIAA), and Princeton University.  Funding was contributed by the FIRST programme from Japanese Cabinet Office, the Ministry of Education, Culture, Sports, Science, and Technology (MEXT), the Japan Society for the Promotion of Science (JSPS),  Japan Science and Technology Agency  (JST),  the Toray Science  Foundation, NAOJ, Kavli IPMU, KEK, ASIAA,  and Princeton University.

Funding for SDSS-III has been provided by the Alfred P. Sloan Foundation, the Participating Institutions, the National Science Foundation, and the U.S. Department of Energy Office of Science. The SDSS-III web site is http://www.sdss3.org/.

SDSS-III is managed by the Astrophysical Research Consortium for the Participating Institutions of the SDSS-III Collaboration including the University of Arizona, the Brazilian Participation Group, Brookhaven National Laboratory, Carnegie Mellon University, University of Florida, the French Participation Group, the German Participation Group, Harvard University, the Instituto de Astrofisica de Canarias, the Michigan State/Notre Dame/JINA Participation Group, Johns Hopkins University, Lawrence Berkeley National Laboratory, Max Planck Institute for Astrophysics, Max Planck Institute for Extraterrestrial Physics, New Mexico State University, New York University, Ohio State University, Pennsylvania State University, University of Portsmouth, Princeton University, the Spanish Participation Group, University of Tokyo, University of Utah, Vanderbilt University, University of Virginia, University of Washington, and Yale University.

\end{acknowledgements}

\bibliographystyle{aa}
\bibliography{references}

\appendix
\onecolumn
\section{Marginalisation over individual lens parameters}\label{sect:appendixa}

In order to sample the posterior probability distribution of the hyper-parameters, $\pr(\hyperp|\data)$, we need to marginalise over the individual object parameters of each lens, evaluating the integrals in \Eref{eq:integral} at each step of the MCMC.
Although $\individ$ is a vector in the eight-dimensional space defined by the parameters listed in \Eref{eq:individ}, some of these parameters are assumed to be known exactly: the lens and source redshift, the S\'{e}rsic index, and the half-light radius of the lens.
As a result, the likelihood term in these variables reduces to a delta function, which effectively transforms \Eref{eq:integral} into an integral over the four-dimensional space in $(\mhalo,\chalo,\mchab,\aimf)$. 

The basic idea for computing the integrals in \Eref{eq:integral} is the following: given a normalised probability distribution $g(\individ)$ and a function $h(\individ)$, we can compute the integral over $\individ$ of the product $g(\individ)h(\individ)$ by drawing a large sample $\left\{\boldsymbol{\psi}^{(k)}\right\}$ from $g(\individ)$ and approximate the integral with the average value of $h(\individ)$ over this sample:
\begin{equation}\label{eq:mcapprox}
\int d\individ h(\individ)g(\individ) \approx \frac{1}{N}\sum_k h(\boldsymbol{\psi}^{(k)}).
\end{equation}

For the purpose of calculating the integral in \Eref{eq:integral}, we could in principle set $g(\individi) = \pr(\individi|\hyperp)$ and $h(\individi) = \pr(\datai|\individi)$. However, drawing samples from $\pr(\individi|\hyperp)$ is complicated by the presence of the lensing cross-section term $\crosssect$, which is not an analytic function of $\individi$.
Instead, we set the sampling distribution to 
\begin{equation}\label{eq:gfunc}
g(\individi) \propto \pcmass(\individi)
\end{equation}
and move the term dependent on the lensing cross-section, the product $A(\hyperp)\crosssect$ in \Eref{eq:hyperdist}, to $h(\individi)$:
\begin{equation}\label{eq:hfunc}
h(\individi) = \pr(\datai|\individi)A(\hyperp)\crosssect(\individi).
\end{equation}
With this choice, $g(\individi)$ is a product of seven Gaussians and a skew Gaussian, $\mathcal{S}$, as summarised in \Eref{eq:cmassterms}.

We draw samples in $(\mhalo,\chalo,\mchab,\aimf)$ from it as follows: given a set of hyper-parameters $\hyperp$, we first draw values of $\mchab$ from $\mathcal{S}(\mchab|\hyperp)$ and values of $\log{\aimf}$ from $\mathcal{A}(\aimf|\hyperp)$, then draw values of $\log{\mhalo}$ from $\mathcal{H}(\mhalo|\mchab,\hyperp)$ and finally draw values of $\log{\chalo}$ from $\mathcal{C}(\chalo|\mhalo)$.
In practice, it is sufficient to sample $\mchab$ from $\mathcal{S}$ only once, since we keep the hyper-parameters describing the stellar mass distribution term fixed. 
For each lens $i$, given a sample $\left\{\boldsymbol{\psi}_i^{(k)}\right\}$ drawn from $g(\individi)$, we calculate the product in \Eref{eq:hfunc} for each point of the sample and use the approximation in \Eref{eq:mcapprox} to calculate the integral in \Eref{eq:integral}.
In order to speed up the evaluation of the likelihood term and the lensing cross-section, we pre-compute, for each lens, the model Einstein radius and $\crosssect$ on a grid of values of stellar mass, halo mass, and halo concentration and obtain the quantities of interest at any point $\individi$ by interpolation.

We calculate the normalisation $A(\hyperp)$ of the probability distribution \Eref{eq:hyperdist} as follows. Since \Eref{eq:hyperdist} must be normalised to unity, $A(\hyperp)$ is defined as
\begin{equation}\label{eq:normfunc}
A(\hyperp)^{-1} = \int d\individ \crosssect(\individ) \pcmass(\individ|\hyperp)
.\end{equation}
This is the lensing cross-section averaged over the distribution $\pcmass(\individ)$, up to a constant that does not depend on the free hyper-parameters. 
We can calculate it by approximating the integral with an average over a sample $\{\boldsymbol{\psi}_i^{(k)}\}$ drawn from $\pcmass(\individi)$.
This time, however, the sample of $\individ$ must span the whole eight-dimensional space defined by \Eref{eq:individ}. Grid interpolation does not help in the evaluation of $\crosssect$, because the high dimensionality of the problem makes it unfeasible.
Instead, we adopt an importance sampling approach. 
We choose a fixed value of the hyper-parameters, $\boldsymbol{\eta}_0$, draw a large sample $\{\boldsymbol{\psi}_i^{(k)}\}$ from $\pcmass(\individ|\boldsymbol{\eta}_0),$ and approximate the integral in \Eref{eq:normfunc} by the mean of $\crosssect$ over the sample, weighted by the ratio between $\pcmass(\individ|\hyperp)$ and $\pcmass(\individ|\boldsymbol{\eta}_0)$:
\begin{equation}
A(\hyperp)^{-1} \approx \frac{1}{N} \sum_k \crosssect(\boldsymbol{\psi}_i^{(k)}) \frac{\pcmass(\boldsymbol{\psi}_i^{(k)}|\hyperp)}{\pcmass(\boldsymbol{\psi}_i^{(k)}|\boldsymbol{\eta}_0)}.
\end{equation}

\end{document}